\documentclass[12pt]{article}

\usepackage{epsfig}
\usepackage{amssymb}
\usepackage{amsbsy}
\usepackage{amsmath}
\usepackage{amsfonts}
\usepackage{mathrsfs} 

\newcommand{\R}{\mathbb{R}}
\newcommand{\C}{\mathbb{C}}
\newcommand{\Z}{\mathbb{Z}}

\newcommand{\bfA}{\mathbf{A}}
\newcommand{\bfE}{\mathbf{E}}
\newcommand{\bfB}{\mathbf{B}}
\newcommand{\bfS}{\mathbf{S}}
\newcommand{\bfL}{\mathbf{L}}
\newcommand{\bfk}{\mathbf{k}}
\newcommand{\bfu}{\mathbf{u}}
\newcommand{\bfx}{\mathbf{x}}
\newcommand{\bfy}{\mathbf{y}}
\newcommand{\bfz}{\mathbf{z}}
\newcommand{\bfp}{\mathbf{p}}
\newcommand{\bfX}{\mathbf{X}}
\newcommand{\bfP}{\mathbf{P}}

\newcommand{\bfa}{\mathbf{a}}
\newcommand{\bfY}{\mathbf{Y}}
\newcommand{\bfZ}{\mathbf{Z}}
\newcommand{\bfJ}{\mathbf{J}}
\newcommand{\bfxi}{\mbox{\boldmath$\xi$}}
\newcommand{\bfPi}{\mbox{\boldmath$\Pi$}}

\newcommand{\bfchi}{\mbox{\boldmath$\chi$}}
\newcommand{\bfbeta}{\mbox{\boldmath$\beta$}}
\newcommand{\bfzeta}{\mbox{\boldmath$\zeta$}}
\newcommand{\bfdel}{\mbox{\boldmath$\nabla$}}
\newcommand{\bfvtheta}{\mbox{\boldmath$\vartheta$}}

\newcommand{\bffX}{\mbox{\boldmath$\mathfrak{X}$}}
\newcommand{\bffY}{\mbox{\boldmath$\mathfrak{Y}$}}
\newcommand{\bfrmx}{{\rm \mathbf{x}}}
\newcommand{\bfrmp}{{\rm \mathbf{p}}}

\newcommand{\fm}{\mathfrak{m}}
\newcommand{\fD}{\mathfrak{D}}
\newcommand{\fU}{\mathfrak{U}}

\newcommand{\bfs}{\mathbf{s}}
\newcommand{\fh}{\mathfrak{h}}
\newcommand{\fK}{\mathfrak{K}}
\newcommand{\fL}{\mathfrak{L}}
\newcommand{\bffK}{\mathbf{\fK}}
\newcommand{\fa}{\mathfrak{a}}
\newcommand{\scrC}{\mathscr{C}}
\newcommand{\fH}{\mathfrak{H}}
\newcommand{\fPa}{\mathscr{P}_{\!\{\fa\}}}

\newcommand{\rh}{{\rm h}}

\newcommand{\be}{\begin{equation}}
\newcommand{\ee}{\end{equation}}
\newcommand{\bea}{\begin{eqnarray}}
\newcommand{\eea}{\end{eqnarray}}
\newcommand{\equa}{&\!\!\!=\!\!\!&}
\newcommand{\equdef}{&\!\!\!:=\!\!\!&}
\newcommand{\nn}{\nonumber}
\newcommand{\kt}{\rangle}
\newcommand{\br}{\langle}

\newcommand{\cum}{\mbox{\scriptsize${\cal M}$}}
\newcommand{\ed}{\end{document}}
\newcommand{\bbr}{\br\!\br}
\newcommand{\kkt}{\kt\!\kt}
\newcommand{\pbr}{\prec\!}
\newcommand{\pkt}{\!\succ}

\newcommand{\cbr}{(\!(}
\newcommand{\ckt}{)\!)}
\newcommand{\f}{\frac}

\newcommand{\p}{\partial}
\newcommand{\ld}{\gamma}
\newcommand{\lnd}{\lambda}
\newcommand{\ep}{\epsilon}
\newcommand{\cH}{\mathcal{H}}
\newcommand{\cK}{\mathcal{K}}
\newcommand{\cF}{\mathcal{F}}
\newcommand{\cG}{\mathcal{G}}
\newcommand{\tcH}{\tilde{\cal H}}
\newcommand{\cHa}{{\cal H}_{\{\fa\}}}
\newcommand{\cUa}{{\cal U}_{\{\fa\}}}
\newcommand{\Ua}{U_{\{\fa\}}}
\newcommand{\suba}{_{\{\fa\}}}

\newcommand{\vrhoa}{\varrho_{_{\{\fa\}}}}

\newcommand{\Aax}{A_{\{\fa\}\,\bfx}}
\newcommand{\Aaxp}{A_{\{\fa\}\,\bfx'}}
\newcommand{\bfAax}{\bfA_{\{\fa\}\,\bfx}}

\newcommand{\xa}{{\bfrmx}_{0\{\fa\}}}
\newcommand{\pa}{{\bfrmp}_{0\{\fa\}}}
\newcommand{\xafm}{{\bfrmx}_{\fm\{\fa\}}}
\newcommand{\pafm}{{\bfrmp}_{\fm\{\fa\}}}
\newcommand{\safm}{s_{\fm\{\fa\}}}
\newcommand{\para}{\ckt_{\{\fa\}}}
\newcommand{\tcK}{\tilde{\cal K}}
\newcommand{\tF}{\tilde{F}}
\newcommand{\tZ}{\tilde{Z}}
\newcommand{\cP}{{\cal P}}
\newcommand{\cPT}{{\cal PT}}
\newcommand{\cCPT}{{\cal CPT}}
\newcommand{\cT}{\mathcal{T}}
\newcommand{\cC}{\mathcal{C}}
\newcommand{\cA}{\mathcal{A}}
\newcommand{\cV}{\mathcal{V}}
\newcommand{\cU}{\mathcal{U}}
\newcommand{\cS}{\mathcal{S}}

\newcommand{\teta}{\tilde{\eta}}
\newcommand{\al}{\alpha}
\newcommand{\trho}{\tilde{\rho}}
\newcommand{\sh}{\mbox{\scriptsize$h$}}

\begin{document}

\title{Quantum Mechanics of Proca Fields}
\author{Farhad~Zamani\thanks{E-mail address: zamani@iasbs.ac.ir} ~and
Ali~Mostafazadeh\thanks{E-mail address:
amostafazadeh@ku.edu.tr} \\
\\
$^*$~Department of Physics, Institute for Advanced Studies in
Basic \\ Sciences, Zanjan 45195-1159, Iran \\
$ ^\dagger$~Department of Mathematics, Ko\c{c} University,
Rumelifeneri Yolu,\\ 34450 Sariyer, Istanbul, Turkey }
\date{ }
\maketitle

\begin{abstract} We construct the most general physically admissible
positive-definite inner product on the space of Proca fields. Up
to a trivial scaling this defines a five-parameter family of
Lorentz invariant inner products that we use to construct a
genuine Hilbert space for the quantum mechanics of Proca fields.
If we identify the generator of time-translations with the
Hamiltonian, we obtain a unitary quantum system that describes
first-quantized Proca fields and does not involve the conventional
restriction to the positive-frequency fields. We provide a rather
comprehensive analysis of this system. In particular, we examine
the conserved current density responsible for the conservation of
the probabilities, explore the global gauge symmetry underlying
the conservation of the probabilities, obtain a probability
current density, construct position, momentum, helicity, spin, and
angular momentum operators, and determine the localized Proca
fields. We also compute the generalized parity ($\cP$),
generalized time-reversal ($\cT$), and generalized charge or
chirality ($\cC$) operators for this system and offer a physical
interpretation for its $\cP\cT$-, $\cC$-, and
$\cC\cP\cT$-symmetries.
\end{abstract}



\section{Introduction}

Quantum Mechanics (QM) and Special Relativity (SR) constitute the
basis for the main body of modern physics developed during the
first half of the twentieth century. The conceptual marriage of QM
and SR is Relativistic Quantum Mechanics (RQM) in which the
one-particle quantum wave functions are identified with the
solutions of an appropriate field equation. The solution space of
these equations provide the representation (Hilbert) spaces for
the irreducible (projective) representations of the Poincar\'e
group that in turn define the elementary particles
\cite{Wigner1,BW,weinberg}. One can construct the Fock space
associated with such a Hilbert space and define the quantum field
operators as operators acting in this space. In this sense, RQM
provides a basis for Quantum Field Theory (QFT). Although a fully
relativistic treatment of elementary particles requires the
methods of QFT, there are processes and energy scales where one
can safely employ RQM. This is for example the case in the
relativistic treatment of quantum systems displaying quantum
superposition and quantum entanglement \cite{Peres-Terno}.

Since the early days of RQM the formulation of a consistent
(first-quantized) quantum theory with a genuine probabilistic
interpretation for bosonic fields has encountered severe
difficulties. It was in an attempt at such a formulation for the
massive scalar (Klein-Gordon) fields that Dirac introduced the
electron's wave equation and discovered the antimatter. His
proposal that led to the development of second quantized field
theories was also mainly motivated by this problem. Recently, we
used the methods of Pseudo-Hermitian QM \cite{M-B} to give a
complete formulation of a genuine quantum mechanical treatment of
Klein-Gordon (KG) fields \cite{MZ-ann06-1}. The purpose of the
present paper is to employ a similar approach to devise a
consistent quantum theory of free first-quantized massive vector
fields.

Research on the theory of massive vector fields started with the
work of Proca \cite{proca-p} whose original aim was to obtain a
description of the four states of electron-positron system using a
four-vector. Proca attempted to pursue the approach of Pauli and
Weisskopf \cite{Pauli-Weisskopf} who had quantized the KG fields
and interpreted the KG conserved current with that of the electric
charge rather than the probability. Although Proca's results did
not serve its original purpose, its mathematical formalism could
be used to treat massive vector fields.

In its manifestly covariant form, Proca's equation for a free
massive vector field $A^\mu$ reads \cite{proca-p}
    \bea
    &&\p_\mu F^{\mu\nu} - \cum^2 A^\nu = 0,
    \label{f-proca}
    \eea
where
    \bea
    &&F^{\mu\nu} := \p^\mu A^\nu - \p^\nu A^\mu,
    \label{F-mu-nu}
    \eea
$\cum:=\f{m c}{\hbar}$ is the inverse of the Compton's wave
length, $m\in\R^+$ is the mass, $\p_\mu
\p^\mu:=\eta^{\mu\nu}\p_\mu\p_\nu$, and
$(\eta^{\mu\nu})=(\eta_{\mu\nu})^{-1}:={\rm diag}(-1,1,1,1)$.

If we apply $\partial_\nu$ to (\ref{f-proca}) and use
$F_{\mu\nu}=-F_{\nu\mu}$, we find that $A^\mu$ satisfies the
Lorentz condition:
    \be
    \p_\mu A^\mu=0.
    \label{Lorentz-gauge-A}
    \ee
Making use of this relation in (\ref{f-proca}) we find
    \be
    (\p_\nu \p^\nu-\cum^2)A^\mu = 0.
    \label{A-proca}
    \ee
This means that the four components of the vector field $A^\mu$
satisfy the KG equation. Note that the Proca equation does not
have a gauge symmetry. Therefore (\ref{Lorentz-gauge-A}) is not a
gauge choice but rather a constraint that is to be imposed on
$A^\mu$. In fact, it is not difficult to show that (\ref{A-proca})
together with (\ref{Lorentz-gauge-A}) are equivalent to the Proca
equation~(\ref{f-proca}). Similarly to the case of KG fields, the
presence of the second time-derivative in (\ref{A-proca}) is
responsible for the difficulties associated with devising a sound
probabilistic interpretation for the Proca fields.

An important motivation for the study of Proca fields is their
close relationship with electromagnetic fields. For example, the
study of Proca fields may shed light on the interesting problem of
the consequences of a nonzero photon rest mass which has been the
subject of intensive research over the past several decades (See
the review article \cite{Tu}.) It might also be possible to
construct a first-quantized quantum theory of a photon by taking
the zero-mass limit of that of a Proca field. This can in
particular led to a resolution of the important issue of the
construction of an appropriate position operator and localized
states for a photon \cite{hawton}.

The literature on the Proca fields is quite extensive. There are a
number of publications that deal with the issue of the consistency
of Proca's theory and the difficulties associated with interacting
massive vector bosons (See \cite{Vij,Silenko} and references
therein.) It turns out that there are various ways of formulating
a relativistic wave equation describing the dynamical states of a
massive vector boson. The famous ones employ the equations of
Proca, Duffin-Kemmer-Petiau (DKP) \cite{KDP}, and
Weinberg-Shay-Good \cite{WSG}. These are equivalent in the absence
of interactions but lead to different predictions upon the
inclusion of interactions.

Among the works that are directly relevant to the subject of the
present article are those of Taketani and Sakata \cite{TS}, Case
\cite{Case}, and Foldy \cite{Foldy}. Using the analogy with the
Maxwell equations, Taketani and Sakata \cite{TS} reduced the
ten-component DKP wave function to a six-component wave function
satisfying a Schr\"{o}dinger equation. They made use of this
representation to study the interaction of the field with an
electromagnetic field. Employing the six-component Taketani-Sakata
(TS) representation, Case \cite{Case} cast the Proca equation into
a form in which the positive- and negative-energy states were
separately described by three-component wave functions. The latter
is the so-called Foldy's canonical representation \cite{Foldy}
which Case used to study the non-relativistic limit of the Proca
equation and obtained the position and spin operators acting on
the TS wave functions. The approaches of Case \cite{Case} and
Foldy \cite{Foldy} involve the use of an indefinite inner product
\cite{indefinite} on the space of six-component TS wave functions.
This seems to be the main reason why these authors did not suggest
any reasonable solution for the problem of the probabilistic
interpretation of their theories. The problem of the construction
of an appropriate position operator and the corresponding
localized states for Proca fields has also been considered by
various authors, e.g.,
\cite{other-position,azc,ne-wi,localizations}. But a universally
accepted solution has not been given. Some other more recent
articles on Proca fields are \cite{proca-recent,Bagrov-Trifonov}.

In \cite{M-CQG,M-ann,M-IJMPA,MZ-ann06-1,MZ-ann06-2}, we have
employed the results obtained within the framework of
Pseudo-Hermitian Quantum Mechanics (PHQM) to formulate a
consistent quantum mechanics of KG fields. Here we pursue a
similar approach to treat the Proca fields. The first step in this
direction has been taken by Jakubsk\'{y} and Smejkal
\cite{Jakubsky} who constructed a one-parameter family of
admissible inner products on the space of Proca fields. In what
follows we will offer a more systematic and general treatment of
this problem. In particular we give a complete characterization of
Lorentz-invariant positive-definite inner products that render the
generator of the time-translations and the helicity operator
self-adjoint. We further construct a position operator and the
corresponding localized states for the Proca fields and examine a
variety of related problems.

In the remainder of this section we briefly review PHQM and give
the notations and conventions that we use throughout the paper.

PHQM \cite{M-B,M-PHQM} has developed in an attempt to give a
mathematically consistent formulation for the $PT$-symmetric
Quantum Mechanics \cite{bender}. In PHQM, a quantum system is
described by a diagonalizable Hamiltonian operator $H$ that acts
in an auxiliary Hilbert space $\cH'$ and has a real spectrum. One
can show (under some general conditions) that these conditions
imply that $H$ is Hermitian with respect to a positive-definite
inner product that is generally different from the defining inner
product $\br\cdot,\cdot\kt$ of $\cH'$, \cite{M-JMP02-2}. This
motivates the following definition of a \emph{pseudo-Hermitian
operator} \cite{M-JMP02-1}: \emph{$H$ is a pseudo-Hermitian
operator  if there is a linear, invertible, Hermitian (metric)
operator $\eta:\cH'\to\cH'$ satisfying $H^\dagger=\eta
H\eta^{-1}$.} This condition is equivalent to the requirement that
$H$ be self-adjoint with respect to the (possibly indefinite)
inner product $\bbr\cdot,\cdot\kkt_\eta:=\br\cdot,\eta\cdot\kt$.
It turns out that for a given pseudo-Hermitian operator $H$, the
metric operator $\eta$ is not unique. If we make a particular
choice for $\eta$, we say that $H$ is
\emph{$\eta$-pseudo-Hermitian}.

A proper subset of pseudo-Hermitian operators is the set of
\emph{quasi-Hermitian} operators \cite{Scholtz-ap}. A
quasi-Hermitian operator $H$ is a pseudo-Hermitian operator that
is $\eta_+$-pseudo-Hermitian for a positive-definite metric
operator $\eta_+$, i.e.,
    \be
    H^\dagger=\eta_+ H \eta_+^{-1}.
    \label{eta-pseudo}
    \ee
This means that $H$ is Hermitian (self-adjoint) with respect to
the positive-definite inner product
    \be
    \bbr\cdot,\cdot\kkt_{\eta_+}:=\br\cdot,\eta_+\cdot\kt.
    \label{pd-inn-prod}
    \ee
In particular $H$ is a diagonalizable operator with a real
spectrum and can be mapped to a Hermitian operator via a
similarity transformation \cite{M-JMP02-2}.

A quasi-Hermitian Hamiltonian operator defines a quantum system
with a consistent probabilistic interpretation provided that one
constructs the physical Hilbert space $\cH$ of the system using
the inner product $\bbr\cdot,\cdot\kkt_{\eta_+}$ and identifies
the observables of the theory with the self-adjoint operators
acting in $\cH$, \cite{Scholtz-ap,M-B}.\footnote{$\cH$ is the
Cauchy completion of the inner product space obtained by endowing
the span of the eigenvector of $H$ (in $\cH'$) with the inner
product $\bbr\cdot,\cdot\kkt_{\eta_+}$.}

Although PHQM employs quasi-Hermitian Hamiltonian operators, the
study of the pseudo-Hermitian Hamiltonians has proved to be
essential in understanding the role of anti-linear symmetries such
as $PT$. As shown in Ref.~\cite{M-JMP03}, given a diagonalizable
pseudo-Hermitian Hamiltonian $H$, one may introduce generalized
parity $(\cP)$, time-reversal $(\cT)$ and charge or chirality
$(\cC)$ operators and establish the $\cC$-, $\cPT$-, and
$\cCPT$-symmetries of $H$ that would respectively generalize $C$,
$PT$, and $CPT$ symmetries \cite{bender-prl} of $PT$-symmetric
quantum mechanics.\footnote{See also \cite{z-ahmed2}.} We recall
that $H$ is $\cP$-pseudo-Hermitian and
$\cT$-anti-pseudo-Hermitian, $\cP$ and $\cT$ are respectively
linear and anti-linear invertible operators, and $\cC$ is a linear
involution (${\cal C}^2=1$) satisfying $\cC=\eta_+^{-1}\cP$,
\cite{M-JMP03}.\footnote{A simple consequence of the latter
relation is that the $CPT$-inner product of \cite{bender-prl} is a
particular example of the inner products~(\ref{pd-inn-prod}) of
\cite{M-JMP02-2,M-JMP02-3,M-NPB02}.}

Next, we give our notations and conventions.

Throughout this paper $a=(a^0,\bfa)$ stands for a four-vector
$a^\mu$, Greek indices take on the values $0,1,2,3$, Latin indices
take on $1,2,3$, and $\varepsilon_{ijk}$ denotes the totally
antisymmetric Levi-Civita symbol with $\varepsilon_{123}=1$. We
employ Einstein's summation convention over repeated indices and
use $\sigma_0$ and $\lnd_0$ to denote the $2\times 2$ and $3\times
3$ identity matrices, respectively. Recall that $\sigma_0$
together with the Pauli matrices
    \be
    \sigma_1=\left(\begin{array}{cc} 0 & 1 \\  1 & 0
    \end{array}\right),~~~~~~~
    \sigma_2=\left(\begin{array}{cc} 0 & -i \\  i & 0
    \end{array}\right),~~~~~~~
    \sigma_3=\left(\begin{array}{cc} 1 & 0 \\  0 & -1
    \end{array}\right),
    \ee
form a basis of the vector space of $2\times 2$ complex matrices,
and that $\{\sigma_0, \sigma_3\}$ is a maximal set of commuting
matrices. We denote their common eigenvectors by $
e_+:=\mbox{\scriptsize$\left(\begin{array}{c}1\\0\end{array}\right)$}$
and
$e_-:=\mbox{\scriptsize$\left(\begin{array}{c}0\\1\end{array}\right)$}$,
where the labels correspond to the eigenvalues of $\sigma_3$.

Similarly, $\lnd_0$ together with
     \bea
    \lnd_1\equa\left(\begin{array}{ccc}
        0 & 1 & 0 \\ 1 & 0 & 0 \\ 0 & 0 & 0
      \end{array}\right),\
      \lnd_2=\left(\begin{array}{ccc}
        0 & -i & 0 \\ i & 0 & 0 \\ 0 & 0 & 0
      \end{array}\right),\
      \lnd_3=\left(\begin{array}{ccc}
        1 & 0 & 0 \\ 0 & -1 & 0 \\ 0 & 0 & 0
      \end{array}\right),\
    \lnd_4=\left(\begin{array}{ccc}
        0 & 0 & 1 \\ 0 & 0 & 0 \\ 1 & 0 & 0
      \end{array}\right),\nn\\
    \lnd_5\equa\left(\begin{array}{ccc}
        0 & 0 & -i \\ 0 & 0 & 0 \\ i & 0 & 0
      \end{array}\right),\
      \lnd_6=\left(\begin{array}{ccc}
        0 & 0 & 0 \\ 0 & 0 & 1 \\ 0 & 1 & 0
      \end{array}\right),\
      \lnd_7=\left(\begin{array}{ccc}
        0 & 0 & 0 \\ 0 & 0 & -i \\ 0 & i & 0
      \end{array}\right),\
    \lnd_8=\f{1}{\sqrt{3}}\left(\begin{array}{ccc}
        1 & 0 & 0 \\ 0 & 1 & 0 \\ 0 & 0 & -2
      \end{array}\right),~~~~
    \eea
form a basis for the vector space of $3\times 3$ complex matrices,
and $\{\lnd_0, \lnd_3, \lnd_8\}$ is a set of maximally commuting
matrices whose common eigenvectors we denote by $
e_{+1}:=\mbox{\scriptsize$\left(\begin{array}{c}1\\0\\0\end{array}\right)$}$,
$e_{-1}:=\mbox{\scriptsize$\left(\begin{array}{c}0\\1\\0\end{array}\right)$}$,
and
$e_{0}:=\mbox{\scriptsize$\left(\begin{array}{c}0\\0\\1\end{array}\right)$}$.
Again the labels $0$ and $\pm1$ are the eigenvalues of $\lnd_3$.

Using the bases $\{\sigma_i\}$ and $\{\lambda_j\}$ we can
construct the basis $\{\Sigma_\fm\}$ for the vector space of
$6\times 6$ complex matrices, where for all $\fm\in\{0, 1, 2,
\cdots,35\}$
    \be
    \Sigma_\fm:=\sigma_i\otimes\lnd_j,~~~~{\rm if}~~~~
    \fm=i+4j.
    \label{Sigma-basis}
    \ee
The operators $\Sigma_0$, $\Sigma_3$, $\Sigma_{12}$,
$\Sigma_{15}$, $\Sigma_{32}$, and $\Sigma_{35}$ form a maximal set
of commuting operators with common eigenvectors
    \be
    e_{\ep,s}:=e_\ep\otimes e_s,
    \label{com-eg-ve-Z}
    \ee
where $\ep\in\{+, -\}$ and $s\in\{-1, 0,+1\}$.

The organization of the paper is as follows. In Section~2, we
present a six-component formulation of the Proca equation,
establish its relation with pseudo-Hermiticity, and construct a
positive-definite metric operator and the corresponding inner
product. In Section~3, we compute the generalized parity $\cP$,
time-reversal $\cT$, and charge or chirality $\cC$ operators and
elaborate on the $\cC$-, $\cPT$-, and $\cCPT$-symmetries of the
Hamiltonian. In Section~4, we derive the expression for the most
general physically admissible positive-definite inner product on
the solution space of the Proca equation and demonstrate the
unitary-equivalence of the representation obtained by the
corresponding Hilbert space and the generator of time-translations
as the Hamiltonian with the Foldy representation. In Section~5, we
obtain and explore the properties of a conserved current density
that supports the conservation of the probabilities. In Section~6,
we introduce a position basis and the associated position wave
functions for the Proca fields. In Section~7, we discuss the
position, spin, helicity, momentum and angular momentum operators,
and construct the relativistic localized states for the Proca
fields. In Section~8, we compute the probability current density
for the spatial localization of a Proca field. In Section~9, we
study the gauge symmetry associated with the conservation of
probabilities. Finally in Section~10, we present a summary of our
main results and discuss the differences between our approach of
finding the most general admissible inner product and that of
Ref.~\cite{Jakubsky}.

\section{Pseudo-Hermiticity and Proca Fields}

\subsection{Covariant Dynamical Field Equation and Helicity States}

To begin our investigation, we briefly formulate the covariant
dynamical theory of Proca fields and review the physical
polarization and helicity states.

Consider a Proca field $A^\mu$ such that for all $x^0\in\R$ and
$\mu\in\{0,1,2,3\}$,
$\int_{\R^3}d^3\bfx\:|A^\mu(x^0,\bfx)|^2<\infty$. Then we can
express (\ref{A-proca}) as a \emph{dynamical equation} in the
Hilbert space $L^2(\R^3)\oplus L^2(\R^3)\oplus L^2(\R^3)\oplus
L^2(\R^3)$, namely
    \be
    \ddot{A}(x^0)+D A(x^0)=0,
    \label{A-d}
    \ee
where a dot denotes a $x^0$-derivative, for all
$\mu\in\{0,1,2,3\}$ and $x^0\in\R$ the functions
$A^\mu(x^0):\R^3\to\C$ are defined by
    \[A^\mu(x^0)(\bfx):=A^\mu(x^0,\bfx),~~~~~~
    \forall \bfx\in\R^3,\]
and $D:L^2(\R^3)\to L^2(\R^3)$ is the Hermitian operator:
    \be
    [D f](\bfx):=[\cum^2-\nabla^2]f(\bfx),~~~~~~
    \forall f\in L^2(\R^3).
    \label{D-def}
    \ee
Clearly, for all $\mu\in\{0,1,2,3\}$ and $x^0\in\R$, $A^\mu(x^0)$
belongs to $L^2(\R^3)$, and $D$ is a positive-definite operator
with eigenvalues $\omega_\bfk^2:=k^2+\cum^2$ and the corresponding
eigenvectors
$\phi_\bfk(\bfx):=(2\pi)^{-3/2}e^{i\bfk\cdot\bfx}=\br\bfx|\bfk\kt$,
where $\bfk\in\R^3$, $k:=|\bfk|$, and $\br\cdot|\cdot\kt$ denotes
the inner product of $L^2(\R^3)$.

It is important to observe that a solution of (\ref{A-d}) is a
Proca field provided that for all $x^0\in\R$ the functions
$A(x^0)$ fulfill the Lorentz condition that we can express as the
constraint:
    \be
    \fL[A(x^0)]:=\dot A^0(x^0)+i\bffK\cdot\bfA(x^0)=0,
    \label{Lorentz-condition-HS}
    \ee
where for all $f\in L^2(\R^3)$, $(\bffK f)(\bfx):=-i\bfdel
f(\bfx)$, and ``$\:\cdot\:$'' is the usual dot product.

The Lorentz condition (\ref{Lorentz-condition-HS}) implies that
there are only three independent components of $A$; using this
condition we can express $A^0(x^0)$ and $\dot A^0(x^0)$ in terms
of $\bfA(x^0)$ and $\dot\bfA(x^0)$. This means that the initial
data for the Proca equation (\ref{A-d}) is given by
$\left(\bfA(x^0_0),\dot\bfA(x^0_0)\right)$ for some initial value
$x_0^0\in\R$ of $x_0$. We shall however employ a manifestly
covariant approach and treat all components of the field $A(x^0)$
on an equal footing. As a result we define the complex vector
space ($\cV$) of solutions of the Proca equation~(\ref{A-d}) as
    \be
    \cV:=\left\{A:=(A^0,\bfA) ~|~\forall
    x^0\in\R,~~
    [\partial_0^2+D]A^\mu(x^0)=0,~\ \fL[A(x^0)]=0 ~\right\}.
    \label{V}
    \ee
It is not difficult to show that the Lorentz condition
(\ref{Lorentz-condition-HS}) may be imposed in the form of a
constraint on the initial data for the dynamical equation
(\ref{A-d}), namely\footnote{The $\ddot A(x^0_0)$ in
$\fL[\dot{A}(x^0_0)]$ stands for $-DA(x^0_0)$.}
    \be
    \fL[A(x^0_0)]=\fL[\dot{A}(x^0_0)]=0.
    \label{constraints}
    \ee
This in turn implies
    \be
    \cV=\left\{A:=(A^0,\bfA) ~|~\forall
    x^0\in\R,~
    [\partial_0^2+D]A^\mu(x^0)=0,~{\rm and}~\exists x_0^0\in\R,~
    \fL[A(x^0_0)]=\fL[\dot{A}(x^0_0)]=0~\right\}.
    \label{V-new}
    \ee

We can express the solution of (\ref{A-d}) in terms of
$(A^\mu(x^0_0),\dot{A}^\mu(x^0_0))$ according to
\cite{M-ann,fulling}
    \be
    A^\mu(x^0)=\cos[(x^0-x^0_0) D^{1/2}]A^\mu(x^0_0)+
    \sin[(x^0-x^0_0) D^{1/2}]D^{-1/2}\dot{A}^\mu(x^0_0).
    \label{g-A-sol}
    \ee
Note that here and throughout this paper we use the spectral
resolution of $D$ to define its powers,
$D^{\alpha}:=\int_{\R^3}d^3\bfk (k^2+\cum^2)^\alpha
|\bfk\kt\br\bfk|$ for all $\alpha\in\R$.

Next, consider the plane wave solutions of (\ref{A-proca}):
    \be
    A^\mu_\ep(\bfk,\sigma;x)=N_{\ep,\bfk}\,a^\mu_\ep(\bfk,\sigma)\,
    e^{-i\ep\omega_\bfk x^0}\phi_\bfk(\bfx),
    \label{plane-wave-z}
    \ee
where $N_{\ep,\bfk}$ are normalization constants (Lorentz
scalars), and $a^\mu_\ep(\bfk,\sigma)$ denotes a set of normalized
four-dimensional polarization vectors fulfilling the Lorentz
condition, i.e.,
    \be
    a_{\ep\,\mu}(\bfk,\sigma)\,a^\mu_\ep(\bfk,\sigma')=
    \delta_{\sigma\sigma'},~~~~~~~~k_\mu\, a^\mu_\ep(\bfk,\sigma)=0.
    \label{a-cond}
    \ee
These relations show that the Proca field (\ref{plane-wave-z}) has
only three physical polarization states ($\sigma\in\{1,2,3\}$). In
a fixed reference frame in which the plane wave has momentum
$\bfk$, we choose to work with a pair of \textit{transverse
polarization vectors}: \cite{weinberg,Greiner}
    \be
    a_\ep(\bfk,1)=(0,\bfa_\ep(\bfk,1)),~~~~~~~
    a_\ep(\bfk,2)=(0,\bfa_\ep(\bfk,2)),
    \label{t-p-v}
    \ee
whose space-like components, $\bfa_\ep(\bfk,\sigma)$ with
$\sigma=1,2$, are normalized vectors perpendicular to $\bfk$. As a
\textit{longitudinal polarization vector} we choose
\cite{weinberg,Greiner}
    \be
    a_\ep(\bfk,3)=
    \left(\f{k}{\cum},\f{k^0}{\cum}\,\f{\bfk}{k}\right),~~~~~~
    k^0=\ep\,\omega_\bfk,
    \label{l-p-v}
    \ee
which together with (\ref{t-p-v}) form an orthonormal set. These
polarization vectors also fulfil the completeness relation
\cite{weinberg,Greiner}:\footnote{The longitudinal polarization
vector used in Ref.~\cite{Jakubsky} is not normalized. Note also
that, unlike the authors of Ref.~\cite{Jakubsky}, we do not fix
the transverse polarization vectors $a_\ep(\bfk,\sigma),
\sigma=1,2$. Although we fix a reference frame in which their
time-like components are zero, their space-like components are
still arbitrary. Instead of fixing these vectors, we only fix the
longitudinal polarization vector (\ref{l-p-v}) and make use of
equations (\ref{a-cond}) and (\ref{pol-com}) to do the necessary
calculations throughout this paper.}
    \be
    \sum_{\sigma=1}^3 a^\mu_\ep(\bfk,\sigma)\,a^\nu_\ep(\bfk,\sigma)=
    \eta^{\mu\nu}+\f{1}{\cum^2}k^\mu k^\nu.
    \label{pol-com}
    \ee

It turns out that the use of the \textit{circular polarization
vectors},
    \be
    u_{\ep,0}(\bfk):=a_\ep(\bfk,3),~~~~~~~
    u_{\ep,\pm1}(\bfk):=\f{1}{\sqrt{2}}\left[a_\ep(\bfk,1)
    \pm i\,a_\ep(\bfk,2)\right],
    \label{cir-pol}
    \ee
simplifies many of the calculations. This is because their spatial
components are eigenvectors of the helicity operator:
    \be
    \fh=\frac{\bffK\cdot\bfS}{|\bffK|}
    =\hat{\bffK}\cdot\bfS=\f{i}{|\bffK|}\left(\begin{array}{ccc}
    0&-\fK_3&\fK_2\\\fK_3&0&-\fK_1\\-\fK_2&\fK_1&0\end{array}\right),
    \label{hel}
    \ee
where $\bffK=:(\fK_1,\fK_2,\fK_3)$, $\hat{\bffK}:=\bffK/|\bffK|$,
and
    \be
    S_1:=\lnd_7,~~~~~~ S_2:=-\lnd_5,~~~~~~
    S_3:=\lnd_2,
    \label{S123}
    \ee
are the angular momentum operators in the spin-one representation
of the rotation group \cite{Greiner-sym}. They satisfy
    \be
    S_i S_j S_k + S_k S_j S_i =
    \delta_{ij} S_k + \delta_{kj} S_i,~~~~~~~~~
    [S_i,S_j]=i\ep_{ijk} S_k.
    \label{S-prop}
    \ee
A simple consequence of these relations is the identity
$\fh^3=\fh$ which we shall make use of in the sequel.

We can also construct the basic (plane wave) solutions of
(\ref{A-proca}) that have a definite helicity. These are given by
    \be
    A^\mu_{\ep,h}(\bfk;x^0,\bfx)=N_{\ep,h}(\bfk)\; u^\mu_{\ep,h}(\bfk)\,
    e^{-i\ep\omega_\bfk x^0}\phi_\bfk(\bfx),~~~~~~~\ep=\pm, \
    \sh=-1,0,1,
    \label{basic-sol}
    \ee
where $N_{\ep,h}(\bfk)$ are normalization constants.

\subsection{Six-Component Formulation and Pseudo-Hermiticity}

In analogy with electromagnetism, we express the antisymmetric
tensor $F^{\mu\nu}$ in terms of the $\bfE$ and $\bfB$ fields
\cite{Greiner-RQM}:
    \be
    F^{ij} =: \varepsilon^{ijm}B_m,~~~~~~~~
    F^{0i} =: E^i.
    \label{eb}
    \ee
Then the Proca equation (\ref{f-proca}) takes the form
    \be
    \begin{array}{ll}
      \bfB=\bfdel\times\bfA, ~~~~~&
      \dot\bfA=-\bfE-\nabla A^0, \\
      A^0=-\cum^{-2} \bfdel\cdot\bfE,~~~~~ &
      \dot\bfE=\cum^2 \bfA+\bfdel\times\bfB.
    \end{array}
    \label{maxeq}
    \ee
Next, we let
$\tcH:=\{\bfY:\R^3\to\C^3|\pbr\bfY,\bfY\pkt\,<\!\infty\}$ denote
the Hilbert space of vector fields, where for all
$\bfY,\bfZ:\R^3\to\C^3$,
$\pbr\bfY,\bfZ\pkt:=\int_{\R^3}d^3\bfx\,\bfY
(\bfx)^*\cdot\bfZ(\bfx)$. If we eliminate $\bfB$ and $A^0$, we can
express (\ref{maxeq}) as a set of dynamical equations in the
Hilbert space $\tcH$, namely
    \be
    \dot\bfA(x^0)=-\bfE(x^0)+\cum^{-2}\fD\bfE(x^0),~~~~~~~
    \dot\bfE(x^0)=D\bfA(x^0)+\fD\bfA(x^0),
    \label{AE}
    \ee
where $\fD:\tcH\to\tcH$ is the Hermitian operator:
    \be
    [\fD\bfY ](\bfx):=\left[[( \bffK\cdot \bfS)^2- \bffK^2]
    \bfY \right](\bfx)=\bfdel(\bfdel\cdot\bfY(\bfx)),
    \label{fD-def}
    \ee
$\bfY\in\tcH$ is arbitrary, and we have used the identities
    \be
    [\bffK^2\bfY](\bfx)=-\nabla^2\bfY(\bfx),~~~~~~~
    \left[( \bffK\cdot \bfS)\bfY\right](\bfx)=
    \nabla\times\bfY(\bfx),
    \label{k2-kdotS}
    \ee
which also imply
    \be
    \bffK\cdot[(\bffK\cdot\bfS)\bfY]=0, ~~~~~~~
    \bffK\cdot[(\bffK\cdot\bfS)^2\bfY]=0.
    \label{div-grad}
    \ee

Next we express (\ref{AE}) as the Schr\"{o}dinger equation (also
known as the Taketani-Sakata equation \cite{TS,Case})
    \be
    i\hbar\dot\Psi(x^0)=H\Psi(x^0),
    \ee
where for all $x^0\in\R$ the state vector $\Psi$ and the
Hamiltonian $H$ are defined by
    \bea
    \Psi(x^0)\equdef\left(\begin{array}{c}
    \bfA(x^0) - i \ld\bfE(x^0) \\
    \bfA(x^0) + i \ld\bfE(x^0) \\
    \end{array}\right),\label{6-comp-psi}\\
    H\equdef\f{\hbar}{2}\,\left(\begin{array}{cc}
    H_1 & H_2 \\
    -H_2 & -H_1 \\
    \end{array}\right),
    \label{6-comp-H}
    \eea
$\ld\in\R-\{0\}$ is an arbitrary constant having the dimension of
length, and $H_1$ and $H_2$ are the following Hermitian operators
that act in $\tcH$.
    \bea
    H_1\equdef\ld(D+\fD)+\ld^{-1}\cum^{-2}(\cum^2-\fD)
    =\ld(\cum^2+(\bffK\cdot\bfS)^2)+\ld^{-1}\cum^{-2}
    (D-(\bffK\cdot\bfS)^2)\,,\label{H1}\\
    H_2\equdef\ld(D+\fD)-\ld^{-1}\cum^{-2}(\cum^2-\fD)
    =\ld (\cum^2+(\bffK\cdot\bfS)^2)-\ld^{-1}\cum^{-2}
    (D-(\bffK\cdot\bfS)^2)\,\label{H2}.
    \eea
Note that one can invert the first equation in (\ref{AE}) (or take
a $x^0$-derivative of the second equation in (\ref{AE}) and use
$\ddot\bfE=-D\bfE$) to obtain
    \be
    \bfE(x^0)=-D^{-1}\left[\cum^2+
    (\bffK\cdot\bfS)^2\right]\dot\bfA(x^0).
    \label{E-in-A}
    \ee
In view of (\ref{6-comp-psi}) and (\ref{E-in-A}), the
six-component state vector $\Psi$ is completely determined in
terms of $\bfA$ and $\dot\bfA$. This means that the space of the
state vectors $\Psi$ is isomorphic (as a vector space) to the
space of the initial conditions $(\bfA(x^0_0),\dot\bfA(x^0_0))$.
Because of the linearity of the Proca equation, the latter is also
isomorphic to the vector space of Proca fields.\footnote{Similarly
to the two-component representation of the KG fields \cite{M-CQG},
the choice of the six-component state vector (\ref{6-comp-psi}) is
not unique. Its general form is $\Psi_{g(x^0)}=g(x^0)\Psi$ where
$g(x^0)\in GL(6,\C)$. (\ref{AE}) is equivalent to the
Schr\"{o}dinger equation
$i\hbar\dot\Psi_{g(x^0)}(x^0)=H_{g(x^0)}\Psi_{g(x^0)}(x^0)$ where
$H_{g(x^0)}:=g(x^0) H g(x^0)^{-1}+i\hbar\dot{g}(x^0)g(x^0)^{-1}$.
The arbitrariness in the choice of $g(x^0)$ is related to a
$GL(6,\C)$ gauge symmetry of the six-component formulation of the
Proca fields. We will take $g(x^0)$ to be the identity matrix.
This is a partial gauge-fixing, because we do not fix $\ld$.
Changing $\ld$ corresponds to a gauge transformation associated
with a $GL(1,\R)$ subgroup of $GL(6,\C)$, namely $
\Psi_\ld\to\Psi_{\ld'}=g(\ld,\ld')\Psi_\ld$ where
$g(\ld,\ld')=\f{1}{2\ld}\mbox{\scriptsize$\left(\begin{array}{cc}
    \ld+\ld' & \ld-\ld' \\
    \ld-\ld' & \ld+\ld' \\
    \end{array}\right)$}\otimes\lnd_0$.
This gauge symmetry has no physical significance, and as we shall
see our final results will not depend on $\ld$.}

The six-component vectors $\Psi(x^0)$ belong to $\C^2\otimes\tcH$.
If we endow the latter with the inner product $\br\cdot,\cdot\kt$
defined by
    \be
    \br\xi,\zeta\kt:=\sum_{i=1}^{2} \pbr\bfxi^i,\bfzeta^i\pkt,
    \label{inn-p-6}
    \ee
for all ${\small\xi=\left(\begin{array}{c}\bfxi^1\\\bfxi^2
\end{array}\right)}$, $\small{\zeta=\left(\begin{array}{c}\bfzeta^1\\
\bfzeta^2\end{array}\right)}\in\C^2\otimes\tcH$, and denote the
corresponding Hilbert space, namely $\tcH\oplus\tcH$, by $\cH'$,
we can view $\Psi(x^0)$ as elements of $\cH'$ and identify $H$
with a linear operator acting in $\cH'$. One can easily check that
$H:\cH'\to\cH'$ is not a Hermitian operator, but it satisfies
$H^{\dagger}=\Sigma_3H\Sigma_3^{-1}$, i.e., it is
$\Sigma_3$-pseudo-Hermitian. This implies that $H$ is self-adjoint
with respect to the inner product
$\bbr\cdot,\cdot\kkt_{\Sigma_3}:=\br\cdot,\Sigma_3\,\cdot\kt$ on
$\C^2\otimes\tcH$. This in turn induces the following inner
product on the space (\ref{V}) of the Proca fields.
    \bea
    \cbr A,A'\ckt_{_{\Sigma_3}}\equdef
    \f{g}{2\ld}\bbr\Psi(x^0),\Psi'(x^0)\kkt_{\Sigma_3}=
    \f{g}{2\ld}\br\Psi(x^0),\Sigma_3\Psi'(x^0)\kt\nn\\
    \equa ig\,\left[\pbr \bfA,(\dot\bfA'+\bfdel A'^0)\pkt-
    \pbr(\dot\bfA+\bfdel A^0),\bfA'\pkt\right],
    \label{s3-inn}
    \eea
where $g\in\R^+$ is a constant. Because of
$\Sigma_3$-pseudo-Hermiticity of $H$, this inner product, which is
sometimes called the \emph{Proca inner product}, is invariant
under the time-evolution generated by $H$, but it is obviously
indefinite.

The dynamical invariance of (\ref{s3-inn}) is associated with a
conserved current density, namely
\cite{Greiner-RQM,Bagrov-Trifonov}
    \be
    J^\mu_{_{\Sigma_3}}(x):=ig \left[{A}_{\nu}(x)^* F^{\nu\mu}(x) -
    {F}^{\nu\mu}(x)^* A_\nu(x)\right].
    \label{s3-current}
    \ee
This is the spin-one analog of the KG current density. We can
express the Proca inner product in a manifestly covariant form in
terms of the current density (\ref{s3-current}). But because this
inner product is indefinite, it cannot be used to make the
solution space of the Proca equation into a genuine inner product
space. Again in analogy to the case of KG fields, one may pursue
the approach of indefinite-metric quantum theories
\cite{indefinite} and restrict to the subspace of positive-energy
solutions (e.g. see \cite{Haller}). But this scheme has the same
difficulties as the corresponding treatment of the KG
fields.\footnote{For example, although the inner product
(\ref{s3-inn}) restricted to the positive-energy solutions is
positive definite, $J^0_{_{\Sigma_3}}(x)$ that should correspond
to the probability density is not generally positive-definite even
for positive-energy fields.}

Next, we will use the positive-definiteness of $D$ to show that
$H$ is a diagonalizable operator with a real spectrum. This
suggests that it is $\eta_+$-pseudo-Hermitian for a
positive-definite metric operator $\eta_+$ \cite{M-JMP02-2}.
Therefore, $H$ is Hermitian with respect to the positive-definite
inner product:
$\bbr\cdot,\cdot\kkt_{\eta_+}:=\br\cdot,\eta_+\cdot\kt$. The
construction of $\eta_+$ requires the solution of the eigenvalue
problem for $H$ and $H^\dagger$.

The eigenvalue problem for the Hamiltonian (\ref{6-comp-H}) may be
easily solved. $H$ has a symmetry generated by
    \be
    \Lambda:=\sigma_0\otimes\fh,
    \label{hilicity-L}
    \ee
which is the helicity operator in the six-component representation
of the Proca fields.\footnote{Hereafter we will omit $\otimes$
wherever there is no risk of confusion.} The simultaneous
eigenvectors of $H$ and $\Lambda$ are given by
    \bea
    \Psi_{\ep,h}(\bfk)\equdef
    \f{1}{2}\left(\begin{array}{c}
    (r_k^{-1}+\ep r_k) \bfu_{\ep,h}(\bfk) -
    \ld r_k^{-1} \bfk u^0_{\ep,h}(\bfk)\\
    (r_k^{-1}-\ep r_k) \bfu_{\ep,h}(\bfk) +
    \ld r_k^{-1} \bfk u^0_{\ep,h}(\bfk)\\
    \end{array}\right) \phi_{_\bfk},
    \label{H-evec}\\
    H\Psi_{\ep,h}(\bfk)\equa E_\ep(\bfk)\Psi_{\ep,h}(\bfk),
    ~~~~~~~~~~~\Lambda\Psi_{\ep,h}(\bfk)=h\,\Psi_{\ep,h}(\bfk),
    \eea
where $\ep\in\{-1,1\}$, $\sh\in\{-1,0,+1\}$, $\bfk\in\R^3$,
$k:=|\bfk|$,
    \[ \omega_\bfk:=\sqrt{k^2+\cum^2},~~~~~
    r_k:=\sqrt{\ld\omega_\bfk},~~~~~
    E_\ep(\bfk):=\ep\hbar\,\omega_\bfk,\]
$u_{\ep,h}(\bfk)$ are circular polarization vectors
(\ref{cir-pol}), and $\phi_{_\bfk}$ are eigenvectors of $D$
corresponding to the eigenvalues $\omega_\bfk$. The eigenvectors
$\Psi_{\ep,h}(\bfk)$ together with
    \be
    \Phi_{\ep,h}(\bfk):=
    \f{1}{2}\left(\begin{array}{c}
    (r_k+\ep r_k^{-1}) \bfu_{\ep,h}(\bfk) -
    \ep \ld r_k^{-1} \bfk u^0_{\ep,h}(\bfk)\\
    (r_k-\ep r_k^{-1}) \bfu_{\ep,h}(\bfk) -
    \ep \ld r_k^{-1} \bfk u^0_{\ep,h}(\bfk)\\
    \end{array}\right) \phi_{_\bfk},
    \label{Hdag-evec}
    \ee
form a complete biorthonormal system for the Hilbert space. This
means that
    \be
    \br\Psi_{\ep,h}(\bfk),\Phi_{\ep',h'}(\bfk')\kt=
    \delta_{\ep,\ep'}\,\delta_{h,h'}\,\delta^3(\bfk-\bfk'),~~~~~~~
    \sum_{\ep=\pm}\sum_{h=0,\pm1} \int_{\R^3} d^3\bfk
    \:|\Psi_{\ep,h}(\bfk)\kt\br\Phi_{\ep,h}(\bfk)| = \Sigma_0,
    \label{psi-phi-biorth}
    \ee
where $\br\cdot,\cdot\kt$ stands for the inner product of $\cH'$,
and for $\xi, \zeta\in\cH'$, $|\xi\kt\br\zeta|$ is the operator
defined by $|\xi\kt\br\zeta|\chi:=\br\zeta,\chi\kt\xi$, for all
$\chi\in\cH'$. Similarly, one can check that indeed
$\Phi_{\ep,h}(\bfk)$ are simultaneous eigenvectors of $H^\dagger$
and $\Lambda^\dagger=\Lambda$ with the same eigenvalues,
$H^\dagger \Phi_{\ep,h}(\bfk)=E_\ep(\bfk)\Phi_{\ep,h}(\bfk)$,
$\Lambda \Phi_{\ep,h}(\bfk)=h\,\Phi_{\ep,h}(\bfk)$, and that $H$
and $\Lambda$ have the following spectral resolutions
    \[H=\sum_{\ep=\pm}\sum_{h=0,\pm1}\int_{\R^3}d^3\bfk\,
    E_\ep(\bfk)\:|\Psi_{\ep,h}(\bfk)\kt\br\Phi_{\ep,h}(\bfk)|,
    ~~~~~
    \Lambda=\sum_{\ep=\pm}\sum_{h=0,\pm1}\int_{\R^3}
    d^3\bfk\:h\:|\Psi_{\ep,h}(\bfk)\kt\br\Phi_{\ep,h}(\bfk)|.\]
Another remarkable property of the biorthonormal system
$\{\Psi_{\ep,h}(\bfk),\Phi_{\ep,h}(\bfk)\}$ is that
    \be
    \br\Psi_{\ep,h}(\bfk),\Psi_{\ep',h'}(\bfk')\kt=
    \br\Psi_{\ep',h'}(\bfk'),\Psi_{\ep,h}(\bfk)\kt=
    \ep\,\ep'\,\br\Phi_{\ep,h}(\bfk),\Phi_{\ep',h'}(\bfk')\kt.
    \label{psi-phi-cond}
    \ee

Using the properties of the polarization vectors, i.e.,
Eqs.~(\ref{a-cond}) and (\ref{pol-com}), we can compute the
positive-definite metric operator, $\eta_+:\cH'\to\cH'$,
associated with the biorthonormal system\\
$\{\Psi_{\ep,h}(\bfk),\Phi_{\ep,h}(\bfk)\}$. The result is
    \be
    \eta_+=\sum_{\ep=\pm}\sum_{h=0,\pm1}\int_{\R^3} d^3\bfk
    |\Phi_{\ep,h}(\bfk)\kt\br\Phi_{\ep,h}(\bfk)|
    =\frac{D^{-1/2}}{2}\,\left(\begin{array}{cc}
     H_1 & H_2 \\
     H_2 & H_1 \\
    \end{array}\right).
    \label{eta+}
    \ee
The inverse of $\eta_+$ has the form
    \be
    \eta_+^{-1}=\sum_{\ep=\pm}\sum_{h=0,\pm1}\int_{\R^3} d^3\bfk
    |\Psi_{\ep,h}(\bfk)\kt\br\Psi_{\ep,h}(\bfk)|=
    \frac{D^{-1/2}}{2}\,\left(\begin{array}{cc}
     H_1 & -H_2 \\      -H_2 & H_1 \\
    \end{array}\right).
    \label{eta+inv}
    \ee

Now we are in a position to compute
$\bbr\cdot,\cdot\kkt_{\eta_+}$. For all $\xi,\zeta\in\cH'$, we let
$\bfxi^i,\bfzeta^i,\bfxi_\pm,\bfzeta_\pm\in\tcH$ be defined by
    \be
    \left(\begin{array}{cc}\bfxi^1\\ \bfxi^2\end{array}\right):=\xi,
    ~~~~~
    \left(\begin{array}{cc}\bfzeta^1\\
    \bfzeta^2\end{array}\right):=\zeta,~~~~~
    \bfxi_\pm:=\bfxi^1\pm\bfxi^2,~~~~
    \bfzeta_\pm:=\bfzeta^1\pm\bfzeta^2.
    \label{xi-zeta}
    \ee
Then in view of (\ref{eta+}), (\ref{H1}), (\ref{H2}), and
(\ref{inn-p-6}),
    \be
    \bbr\xi,\zeta\kkt_{\eta_+}=\f{1}{2}\,
    \left[\ld\pbr\bfxi_+,D^{-1/2}
    [\cum^2+(\bffK\cdot\bfS)^2]\bfzeta_+\pkt+\f{1}{\ld\cum^2}
    \pbr\bfxi_-,D^{-1/2}[D-(\bffK\cdot\bfS)^2]\bfzeta_-\pkt\right].
    \label{inn+}
    \ee
If we view $\cH'$ as a complex vector space and endow it with the
inner product (\ref{inn+}), we obtain a new inner product space
whose Cauchy completion yields a Hilbert space which we denote by
$\cK$.

Next, for all $x^0\in\R$ we define $U_{x^0}:{\cal V}\to{\cal H}'$
according to
    \be
    U_{x^0}A:=\f{1}{2}\,\sqrt{\f{\kappa}{\ld\cum}}\:\Psi(x^0),~~~~~~
    \forall A\in{\cal V},
    \label{U-zero}
    \ee
where $\kappa\in\R^+$ is a fixed but arbitrary constant. We can
use $U_{x^0}$ to endow the complex vector space ${\cal V}$ of
Proca fields with the positive-definite inner product
    \be
    \cbr A,A'\ckt:=\bbr U_{x^0}A,U_{x^0}A'\kkt_{\eta_+}
    =\f{\kappa}{4\ld\cum} \bbr \Psi(x^0),\Psi'(x^0)\kkt_{\eta_+}.
    \label{inn-inv-1}
    \ee
Because $\eta_+$ does not depend on $x^0$, the inner product
$\bbr\cdot,\cdot\kkt_{\eta_+}$ is invariant under the dynamics
generated by $H$, \cite{M-JMP02-1}. This in turn implies that the
right-hand side of (\ref{inn-inv-1}) should be $x^0$-independent.
In order to see this, we substitute (\ref{6-comp-psi}) and
(\ref{inn+}) in (\ref{inn-inv-1}) and use (\ref{AE}) and
(\ref{fD-def}) to derive
    \bea
    \cbr A,A'\ckt\equa\f{\kappa}{2\cum}
    \left[\pbr\bfA(x^0),D^{-1/2}[\cum^2+(\bffK\cdot\bfS)^2]\bfA'(x^0)\pkt
    \right.\nn\\
    &&\vspace{2cm}\left.+\cum^{-2}\!\pbr(\dot\bfA(x^0)+
    i\bffK\,A^0(x^0)),
    D^{-1/2}[D-(\bffK\cdot\bfS)^2](\dot\bfA'(x^0)+i\bffK\,
    A'^0(x^0))\pkt\right] \nn\\
    \equa\f{\kappa}{2\cum}\left[\pbr\bfA(x^0),D^{-1/2}\dot\bfE'(x^0)\pkt-
    \pbr\bfE(x^0),D^{-1/2}\dot\bfA'(x^0)\pkt\right],
    \label{inn-inv-2}
    \eea
where $\bfE(x^0)=-\dot\bfA(x^0)-i\bffK\, A^0(x^0)$. We can use
(\ref{A-d}) or equivalently (\ref{AE}) to check that the
$x^0$-derivative of the right-hand side of (\ref{inn-inv-2})
vanishes identically. Therefore, (\ref{inn-inv-2}) provides a
well-defined inner product on ${\cal V}$. Endowing ${\cal V}$ with
this inner product and (Cauchy) completing the resulting inner
product space we obtain a separable Hilbert space which we shall
denote by ${\cal H}$. This is the physical Hilbert space of the
relativistic quantum mechanics of the Proca fields.

The inner product~(\ref{inn-inv-2}) is identical with an inner
product obtained in \cite{Jakubsky}. We will show in Section~4
that it is a special example of a larger class of invariant inner
products and that it has the following appealing properties:
    \begin{enumerate}
    \item  It is not only positive-definite but relativistically
invariant.\footnote{We give a manifestly covariant expression for
this inner product in Section 5.}
    \item Its restriction to the
subspace of positive-frequency Proca fields coincides with the
restriction of the indefinite Proca inner product (\ref{s3-inn})
to this subspace.
    \end{enumerate}

As seen from (\ref{inn-inv-1}), the operator $U_{x^0}$ for any
value of $x^0\in\R$ is a unitary operator mapping ${\cal H}$ to
${\cal K}$. Following \cite{M-ann,M-IJMPA} we can use this unitary
operator to define a Hamiltonian operator $\rh$ acting in ${\cal
H}$ that is unitary-equivalent to $H$. Let $x^0_0\in\R$ be an
arbitrary initial $x^0$, and $\rh:{\cal H}\to{\cal H}$ be defined
by
    \be
    \rh:=U_{x^0_0}^{-1}\,H\,U_{x^0_0}.
    \label{h}
    \ee
Then, using (\ref{maxeq}), (\ref{6-comp-psi}) -- (\ref{H2}) and
(\ref{U-zero}), we can easily show that for any $A\in\cV$,
    \be
    \rh A=i\hbar\dot A,
    \label{h=}
    \ee
where $\dot A$ is the element of ${\cal V}$ defined by: $\dot
A(x^0):=\frac{d}{dx^0}\,A(x^0)$, for all $x^0\in\R$. As discussed
in \cite{M-ann} for the case of KG fields, one must not confuse
(\ref{h=}) with a time-dependent Schr\"odinger equation giving the
$x^0$-dependence of $A(x^0)$. This equation actually defines the
action of the operator $\rh$ on the field $A$. $\rh$ generates a
time-evolution, through the Schr\"odinger equation
    \be
    i\hbar\frac{d}{dx^0}A_{x^0}=\rh A_{x^0},
    \label{sch-eq-h}
    \ee
that coincides with a time-translation in the space ${\cal V}$ of
the Proca fields; if $A_0=A_{x^0_0}$ is the initial value for the
one-parameter family of elements $A_{x^0}$ of ${\cal V}$, then for
all $x^0,x^{0'}\in\R$, $A_{x^0}(x^{0'})=
(e^{-i(x^0-x^0_0)\rh/\hbar}A_0)(x^{0'})=A_0(x^{0'}+x^0-x^0_0)$.
Furthermore, using the fact that $U_{x^0_0}$ is a unitary operator
and that $H$ is Hermitian with respect to the inner product
(\ref{inn+}) of ${\cal K}$, we can infer that $\rh$ is Hermitian
with respect to the inner product~(\ref{inn-inv-1}) of ${\cal H}$.
This shows that time-translations correspond to unitary
transformations of the physical Hilbert space ${\cal H}$.

Next, we define ${\cal U}:{\cal H}\to{\cal H}'$ and $H':{\cal
H}'\to{\cal H}'$ by
    \bea
    {\cal U}\equdef\rho\: U_{x^0_0}
    \label{U=}\\
    H'\equdef{\cal U}\: \rh \:{\cal U}^{-1}=\rho H\rho^{-1},
    \label{H-prime}
    \eea
where $\rho$ is the unique positive square root of $\eta_+$, i.e.,
$\rho:=\sqrt\eta_+$. It is not difficult to see that
    \be
    \rho=\f{1}{2\cum\sqrt{\ld}}\,\left(\begin{array}{cc}
    \rho_+ & \rho_-\\ \rho_- & \rho_+\end{array}\right),~~~~~~
    \rho^{-1}=\f{1}{2\cum\sqrt{\ld}}\,\left(\begin{array}{cc}
    \rho_+ & -\rho_-\\
    -\rho_- & \rho_+\end{array}\right),
    \label{rho}
    \ee
where $\rho_\pm:\tcH\to\tcH$ are Hermitian operators given by
    \be
    \rho_\pm:=(\ld\cum\mp 1)[D^{1/4}-\cum D^{-1/4}]\,\fh^2+
    \ld\cum^2 D^{-1/4}\pm D^{1/4},
    \label{rho-ij}
    \ee
and $\fh$ is the helicity operator (\ref{hel}). We can check that
the operator $\rho$ viewed as mapping ${\cal K}$ to ${\cal H'}$ is
a unitary transformation; using $\rho^\dagger=\rho=\sqrt\eta_+$,
we have $\br\rho\xi,\rho\zeta\kt=\bbr\xi,\zeta\kkt_{\eta_+}$ for
all $\xi,\zeta\in{\cal K}$. This in turn implies that ${\cal
U}:{\cal H}\to{\cal H}'$ is also a unitary transformation,
    \be
    \br{\cal U}A,{\cal U}A'\kt=\cbr A,A'\ckt,
    ~~~~~~~~~~~\forall A,A'\in{\cal H},
    \label{unitary-u}
    \ee
and that $H'$ must be a Hermitian Hamiltonian operator acting in
${\cal H}'$.

We can compute $H'$ by substituting (\ref{rho}), (\ref{rho-ij})
and (\ref{6-comp-H}) -- (\ref{H2}) in (\ref{H-prime}). This yields
    \be
    H'=\hbar\,\left(\begin{array}{cc}
    \sqrt D & 0\\
    0 & -\sqrt D \end{array}\right)=\hbar\sqrt D\,\Sigma_3,
    \label{foldy-H}
    \end{equation}
which is manifestly Hermitian with respect to the inner product
$\br\cdot,\cdot\kt$ of ${\cal H}'$. The Hamiltonian $H'$ is
precisely the Foldy Hamiltonian~\cite{Foldy,Case}. Here we
obtained it by a systematic application of the methods of PHQM
\cite{M-B}.

Next, we derive the explicit form of ${\cal U}$ and its inverse.
Using (\ref{U=}), (\ref{rho}), (\ref{rho-ij}), (\ref{U-zero}), and
(\ref{6-comp-psi}), we have for all $A\in{\cal H}$,
\footnote{Transformation (\ref{u-A=}) is known as the Foldy's
transformation, although it differs slightly from the expression
given by Foldy \cite{Foldy} and Case \cite{Case}, namely
$\cU_{_{\rm Foldy}}:=\Sigma_3\,\cU$. This difference turns out not
to have any effect on the definition of the physical observables
such as the position operator.}
    \be
    \cU A=\frac{1}{2}\sqrt{\f{\kappa}{\cum}}\,
    \left(\begin{array}{c}
    \fU \bfA(x^0_0)-i\,\fU^{-1}\bfE(x^0_0)\\
    \fU\bfA(x^0_0)+i\,\fU^{-1}\bfE(x^0_0)
    \end{array}\right),
    \label{u-A=}
    \ee
where the operator $\fU:\tcH\to\tcH$ and its inverse are given
by\footnote{It is interesting to see that the arbitrary parameter
$\ld$, introduced in the six-component formulation of the Proca
equation, does not appear in (\ref{foldy-H}) and (\ref{u-A=}).}
    \be
    \fU=[D^{1/4}-\cum D^{-1/4}] \fh^2+\cum
    D^{-1/4},~~~~~~~
    \fU^{-1}=[D^{-1/4}-\cum^{-1} D^{1/4}] \fh^2+\cum^{-1}D^{1/4}.
    \ee
The inverse ${\cal U}^{-1}$ of ${\cal U}$ is also easy to
calculate. Let $\xi\in{\cal H}'$ be a six-component vector (as in
(\ref{xi-zeta})) with components $\bfxi^1$ and $\bfxi^2$. Then in
view of (\ref{u-A=}), ${\cal U}^{-1}\xi$ is the Proca field
$A\in\cH$ satisfying the following initial conditions:
    \bea
    A^0(x^0_0)&=& \f{D^{-1/4}}{\sqrt{\cum\kappa}}\,
    \bffK\cdot(\bfxi^1-\bfxi^2),
    ~~~~~~~~~~~
    \dot A^0(x^0_0)=-i\,\f{D^{1/4}}{\sqrt{\cum\kappa}}\,
    \bffK\cdot(\bfxi^1+\bfxi^2),\label{U-inverse-0}\\
    \bfA(x^0_0)&=&\sqrt{\f{\cum}{\kappa}}\,\fU^{-1}\,(\bfxi^1+\bfxi^2),
    ~~~~~~~~~~~
    \dot\bfA(x^0_0)=-i\,\sqrt{\f{\cum}{\kappa}}\,
    D^{1/2}\,\fU^{-1}\,(\bfxi^1-\bfxi^2),
    \label{U-inverse-i}
    \eea
where we have made use of equations (\ref{constraints}),
(\ref{AE}), (\ref{div-grad}), (\ref{6-comp-psi}), (\ref{U-zero}),
(\ref{rho}), and (\ref{rho-ij}). By virtue of (\ref{g-A-sol}), for
all $x^0\in\R$, we have
    \bea
    A^0(x^0)&=& \f{D^{-1/4}}{\sqrt{\cum\kappa}}\,
    \left[e^{-i(x^0-x^0_0) D^{1/2}}\bffK\cdot\bfxi^1-
    e^{i(x^0-x^0_0) D^{1/2}}\bffK\cdot\bfxi^2\right],\\
    \bfA(x^0)&=&\sqrt{\f{\cum}{\kappa}}\,\fU^{-1}\,
    \left[e^{-i(x^0-x^0_0) D^{1/2}}\bfxi^1+
    e^{i(x^0-x^0_0) D^{1/2}}\bfxi^2\right].
    \eea
In complete analogy with the case of KG fields \cite{M-IJMPA}, we
find that the pairs $({\cal H},\rh)$, $({\cal K},H)$, and $({\cal
H}',H')$ are unitarily equivalent. Therefore they represent the
same quantum system.

\section{$\cPT$, $\cC$ and $\cCPT$-symmetries of Proca
Fields}\label{CPT-section}

According to \cite{M-JMP03}, the generalized parity ($\cP$),
time-reversal ($\cT$), and charge grading or chirality ($\cC$)
operators for the six-component Proca fields are given by
    \bea
    \cP \equdef\sum_{\ep=\pm}\sum_{\sigma=1}^3\int_{\R^3}
    d^3\bfk~\sigma_{\ep,\sigma}(\bfk)\,|\Phi_{\ep,\sigma}(\bfk)\kt
    \br\Phi_{\ep,\sigma}(\bfk)|,
    \label{Ps-Proca}\\
    \cT \equdef\sum_{\ep=\pm}\sum_{\sigma=1}^3\int_{\R^3}d^3\bfk~
    \sigma_{\ep,\sigma}(\bfk)\,|\Phi_{\ep,\sigma}(\bfk)\kt\,\star\,
    \br\Phi_{\ep,\sigma}(\bfk)|,
    \label{Ts-Proca}\\
    \cC \equdef\sum_{\ep=\pm}\sum_{\sigma=1}^3\int_{\R^3}d^3\bfk~
    \sigma_{\ep,\sigma}(\bfk)\,
    |\Psi_{\ep,\sigma}(\bfk)\kt\br\Phi_{\ep,\sigma}(\bfk)|,
    \label{Cs-Proca}
    \eea
where $\sigma_{\ep,\sigma}(\bfk)$ are arbitrary signs ($\pm$), and
$\star$ is the complex-conjugation operator defined, for all
complex numbers $z$ and state vectors $\Psi,\Phi$, by $\star
z:=z^*$ and $(\star\br\Phi|)|\Psi\kt
:=\star\,\br\Phi|\Psi\kt=\br\Psi|\Phi\kt$. Note that we use the
biorthonormal basis
$\{\Psi_{\ep,\sigma}(\bfk),\Phi_{\ep,\sigma}(\bfk)\}$ to define
$\cP$, $\cT$, and $\cC$. This basis can be obtained by replacing
the circular polarization vectors $u_{\ep,h}(\bfk)$ by the linear
polarization vectors $a_{\ep}(\bfk,\sigma)$ in (\ref{H-evec}) and
(\ref{Hdag-evec}).

Clearly there is an infinity of choices for the signs
$\sigma_{\ep,\sigma}(\bfk)$ each leading to a different $\cP$,
$\cT$, and $\cC$. Following \cite{M-IJMPA} we will adopt the
natural choice associated with the label $\epsilon$ appearing in
the expression for the eigenvalues and eigenvectors of $H$. This
yields
    \bea
    \cP\equdef\sum_{\ep=\pm}\sum_{h=0,\pm1}\int_{\R^3}
    d^3\bfk~\ep\,|\Phi_{\ep,h}(\bfk)\kt\br\Phi_{\ep,h}(\bfk)|,
    \label{P-Proca}\\
    \cT\equdef\sum_{\ep=\pm}\sum_{h=0,\pm1}\int_{\R^3}d^3\bfk~\ep\,
    |\Phi_{\ep,h}(\bfk)\kt\:\star\:\br\Phi_{\ep,-h}(\bfk)|,
    \label{T-Proca}\\
    \cC\equdef\sum_{\ep=\pm}\sum_{h=0,\pm1}\int_{\R^3}d^3\bfk~\ep\,
    |\Psi_{\ep,h}(\bfk)\kt\br\Phi_{\ep,h}(\bfk)|.
    \label{C-Proca}
    \eea
The consequences of these relations are identical with those
obtained in \cite{M-IJMPA} for the KG fields. Because of the close
analogy with the case of KG fields, here we omit the details and
give a summary of the relevant results.

Substituting (\ref{H-evec}) and (\ref{Hdag-evec}) in
(\ref{P-Proca}) -- (\ref{C-Proca}) yields
    \be
    \cP=\Sigma_3,~~~~~~~~
    \cT=\Sigma_3\:\star,~~~~~~~~
    \cPT=\star.
    \label{PT}
    \ee
Hence, the $\cPT$-symmetry of the Hamiltonian (\ref{6-comp-H})
means that it is a real operator. Note that $\cP^2=\cT^2=\Sigma_0$
and $\cPT=\cP\cdot\cT$, which is a direct consequence of
(\ref{psi-phi-cond}), \cite{M-JMP03}. Similarly, we find
    \be
    \cC=\f{D^{-1/2}}{2}\,\left(\begin{array}{cc}
     H_1 & H_2 \\
     -H_2 & -H_1 \\
    \end{array}\right)=\hbar^{-1}D^{-1/2}H=\frac{H}{\sqrt{H^2}},
    \label{C-Proca=H}
    \end{equation}
which in view of (\ref{eta+}) is consistent with the identity
$\cC=\eta_+^{-1}\cP$ (equivalently $\eta_+={\cal PC}$
\footnote{Note that $\cP$, $\cT$ and the metric operator $\eta_+$
depend on the choice of the biorthonormal system, while $\cC$ is
independent. In \cite{ZG} this fact has not been considered in the
factorization of the metric operator.}) \cite{M-JMP03}. According
to (\ref{C-Proca=H}), $\cC$ is a $\Z_2$-grading operator for the
Hilbert space that splits it into the spans of the eigenvectors of
$H$ with positive and negative eigenvalues, respectively.

We can use the unitary operator $\rho:\cK\to\cH'$ to express the
generalized parity, time-reversal, and chirality operators in the
Foldy representation, namely $\cP':=\rho\,\cP\rho^{-1}$,
$\cT':=\rho\,\cT\rho^{-1}$, and $\cC':=\rho\,\cC\rho^{-1}$. The
symmetry generators $\cP'\cT'$ and $\cC'$ that commute with the
Foldy Hamiltonian $H'$ have the form
    \be
    \cP'\cT'=\cPT=\star,~~~~~~~~~~\cC'=
    \frac{H'}{\sqrt{H^{'2}}}=\Sigma_3.
    \label{PT-prime}
    \ee
Clearly, the $\cP'\cT'$-symmetry of $H'$ is related to the fact
that $H'$ is a real operator, and the $\cC'$-symmetry of $H'$
arises because it is proportional to $\cC'$.

Similarly we can use the unitary operator $\cU:\cH\to\cH'$ to
define the generalized parity, time-reversal, and chirality
operators for the ordinary Proca fields $A\in\cH$. This yields
    \be
    {\rm P}:=\cU^{-1}{\cal P}'\cU,~~~~~~~~~~
    {\rm T}:=\cU^{-1}{\cal T}'\cU,~~~~~~~~~~
    {\rm C}:=\cU^{-1}{\cal C}'\cU.
    \label{P-T-C}
    \ee
Using Eqs.~(\ref{u-A=}) -- (\ref{U-inverse-i}), and
(\ref{g-A-sol}), we may obtain explicit expressions for ${\rm P}$
and ${\rm T}$. Here we give the corresponding expressions for
${\rm PT}$ and ${\rm C}$, that actually generate symmetries of the
Hamiltonian $\rh$. The result is, for all $A\in\cH$ and
$x^0\in\R$,
    \bea
    [{\rm PT}A](x^0)&=&(-A^0(-x^0)^*,\bfA(-x^0)^*),
    \label{time-rev}\\
    \left[{\rm C}A\right](x^0)&=&i D^{-1/2}\dot A(x^0)=:A_c(x^0).
    \label{A-c}
    \eea
These relations imply that, similarly to the case of KG field
\cite{M-IJMPA}, ${\rm PT}$ is just the ordinary time-reversal
operator \cite{Foldy}, and ``{\em the ${\rm PT}$-symmetry of $\rh$
means that the order in which one performs a time-translation and
a time-reversal transformation on a Proca field is not
important}.'' Furthermore, chirality operator acting in $\cH$ is
simply given by ${\rm C}=i D^{-1/2}\p_0$, which is a Lorentz
scalar \cite{MZ-ann06-1}. Recalling ${\rm C}^2 A=A$, we observe
that ${\rm C}:\cV\to\cV$ is an involution, ${\rm C}^2=1$.
Therefore, we can use it to split $\cV$ into the subspaces
$\cV_\pm$ of $\pm$-energy Proca fields according to
$\cV_\pm:=\{A_\pm\in\cV | {\rm C} A_\pm = \pm A_\pm\}$. Clearly,
for any $A\in\cV$, we can use ${\rm C}$ to introduce the
corresponding $\pm$-energy components:
    \be
    A_\pm:=\frac{1}{2}\,(A\pm {\rm C} A)\in \cV_\pm.
    \label{A-pm-field}
    \ee
Clearly $A=A_++A_-$, so $\cV=\cV_+\oplus\cV_-$. Restricting the
inner product (\ref{inn-inv-2}) to $\cV_\pm$ (and Cauchy
completing the resulting inner product spaces) we obtain Hilbert
subspaces $\cH_\pm$ of $\cH$.

Next we recall that $\cC'$ is a Hermitian operator acting in
$\cH'$. Thus, in view of (\ref{P-T-C}), and (\ref{unitary-u}), we
have $\cbr A,{\rm C} A'\ckt=\cbr{\rm C} A,A'\ckt$. Therefore,
${\rm C}:\cH\to\cH$ is a Hermitian involution\footnote{This marks
its similarity to the chirality operator $\gamma^5$ for spin 1/2
fields.} and for all $A_\pm\in\cV_\pm$, $\cbr A_+,A_-\ckt=0$. This
in turns implies $\cH=\cH_+\oplus\cH_-$. The generalized chirality
operator ${\rm C}$ is actually the grading operator associated
with this orthogonal direct sum decomposition of $\cH$. In other
words, similarly to the case of KG fields \cite{M-IJMPA}, ${\rm
C}$ is a Hermitian involution that decomposes the Hilbert space
into its $\pm$-energy subspaces. As a result, ``{\em the ${\rm
C}$-symmetry of $\rh$ means that the energy of a free Proca field
does not change sign under time-translations}.''

\section{The Most General Admissible Inner
Product}\label{gen-metric-sec}

The metric operator $\eta_+$ and the corresponding invariant
positive-definite inner product depend on the choice of the
biorthonormal system $\{\Psi_{\ep,h}(\bfk),\Phi_{\ep,h}(\bfk)\}$.
The most general invariant positive-definite inner product
corresponds, therefore, to the most general biorthonormal system
that consists of the eigenvectors of $H$ and $H^\dagger$. In the
construction of $\{\Psi_{\ep,h}(\bfk),\Phi_{\ep,h}(\bfk)\}$ we
have chosen a set of eigenvectors of $H$ that also diagonalize the
helicity operator $\Lambda$. As a result, both $H$ and $\Lambda$
are $\eta_+$-pseudo-Hermitian. This in turn is equivalent to the
condition that $H$ and $\Lambda$ are among the physical
observables of the system. In the following we construct the most
general Lorentz invariant positive-definite inner product that
fulfills this condition.\footnote{This amounts to restricting the
choice of the metric operator in the spirit of \cite{Scholtz-ap}.
It does not however fix the metric operator (up to a trivial
scaling) because $H$ and $\Lambda$ do not form an irreducible set
of operators.}

The most general positive-definite operator $\teta_+$ that renders
both $H$ and $\Lambda$~ $\teta_+$-pseudo-Hermitian has the form
$\teta_+:=\cA^\dagger \eta_+ \cA$ where $\cA$ is an invertible
linear operator commuting with $H$ and $\Lambda$
\cite{M-NPB02,M-JMP03}. We can express $\cA$ as
    \be
    \cA=\sum_{\ep=\pm}\sum_{h=0,\pm1}\int_{\R^3} d^3\bfk\:
    \al_{\ep,h}(\bfk)|\Psi_{\ep,h}(\bfk)\kt
    \br\Phi_{\ep,h}(\bfk)|,
    \label{cA-g}
    \ee
where $\al_{\ep,h}(\bfk)$ are arbitrary nonzero complex
numbers.\footnote{The condition that $H$ is
$\teta_+$-pseudo-Hermitian yields
$\cA=\sum_{\ep=\pm}\sum_{h=0,\pm1}\sum_{h'=0,\pm1}\int_{\R^3}
d^3\bfk\:
\al^\ep_{h,h'}(\bfk)|\Psi_{\ep,h}(\bfk)\kt\br\Phi_{\ep,h'}(\bfk)|$
which does not commute with $\Lambda$ unless
$\al^\ep_{h,h'}=\delta_{h,h'}\al_{\ep,h}$.} This implies
    \be
    \teta_+:=\cA^\dagger \eta_+ \cA=\sum_{\ep=\pm}\sum_{h=0,\pm1}
    \int_{\R^3} d^3\bfk\: |\al_{\ep,h}(\bfk)|^2
    \:|\Phi_{\ep,h}(\bfk)\kt\br\Phi_{\ep,h}(\bfk)|.
    \label{eta-+g}
    \ee
Substituting (\ref{H-evec}) and (\ref{Hdag-evec}) in
(\ref{eta-+g}), and carrying out the necessary calculations, we
find
    \be
    \teta_+=\left[L^+_+ M+(L^+_--L^0_-) \sigma_3-L^0_+
    N\right]\fh^2+\left[L^-_+ M +
    L^-_-\sigma_3\right]\fh + L^0_+ N + L^0_-\sigma_3,
    \label{eta+g}
    \ee
where
    \bea
    L^\ep_\pm\equdef\f{1}{4}\int d^3\bfk
    \left[|\al_{+,1}(\bfk)|^2+\ep|\al_{+,-1}(\bfk)|^2\pm
    |\al_{-,1}(\bfk)|^2\pm\ep|\al_{-,-1}(\bfk)|^2\right]
    |\bfk\kt\br\bfk|,\label{L-ep-pm}\\
    L^0_\pm\equdef\f{1}{2}\int d^3\bfk
    \left[|\al_{+,0}(\bfk)|^2\pm|\al_{-,0}(\bfk)|^2\right]
    |\bfk\kt\br\bfk|,\label{L-0-pm}\\
    M\equdef\f{D^{-1/2}}{2\ld}\,\left(\begin{array}{cc}
      \ld^2 D+1 & \ld^2 D-1 \\
      \ld^2 D-1 & \ld^2 D+1 \\
    \end{array}\right), \\
    N\equdef\f{D^{-1/2}}{2\ld\cum^2}\,\left(\begin{array}{cc}
      \ld^2\cum^4+D & \ld^2\cum^4-D \\
      \ld^2\cum^4-D & \ld^2\cum^4+D \\
    \end{array}\right).\label{N-matrix}
    \eea

Having computed $\teta_+$ we can easily obtain the corresponding
inner product according to
$\bbr\cdot,\cdot\kkt_{\teta_+}:=\br\cdot,\teta_+\cdot\kt$. This
yields, for all $\xi,\zeta\in\cH'$,
    {\small\be
    \bbr\xi,\zeta\kkt_{\teta_+} =
    \f{1}{2}\left[\ld\!\pbr\bfxi_+,\Theta_{+,+1}\bfzeta_+\pkt
    +\ld^{-1}\!\!\pbr\bfxi_-,\Theta_{+,-1}\bfzeta_-\pkt\right]
    +\pbr\bfxi^1,\Theta_{-,0}\bfzeta^1\pkt-
    \pbr\bfxi^2,\Theta_{-,0}\bfzeta^2\pkt,
    \label{inn+g}
    \ee}
where the operators $\Theta_{\ep,h}:\tcH\to\tcH$ are defined by
    \be
    \Theta_{\ep,h}:=[L^+_\ep D^{h/2}-
    \cum^{2h} L^0_\ep D^{-h/2}]\fh^2
    +L^-_\ep D^{h/2}\fh+\cum^{2h} L^0_\ep D^{-h/2}.
    \label{Theta-eph}
    \ee
If we view $\cH'$ as a complex vector space and endow it with this
inner product, we obtain a new inner product space whose Cauchy
completion yields a Hilbert space which we denote by $\tcK$.
Finally, we recall that, defining $\cbr A,A'\para:=\bbr
U_{x^0}A,U_{x^0}A'\kkt_{\teta_+}$, we can endow the space $\cV$ of
the solutions of Proca equation with the positive-definite inner
product
    \bea
    \cbr A,A'\para\equdef\f{\kappa}{2\cum}
    \left[\pbr\bfA(x^0),\Theta_{+,+1}\bfA'(x^0)\pkt+
    \pbr\bfE(x^0),\Theta_{+,-1}\bfE'(x^0)\pkt
    \right.\nn\\
    & &\vspace{2cm}\hspace{1.25cm}\left.+i\left\{
    \pbr\bfE(x^0),\Theta_{-,0}\bfA'(x^0)\pkt-
    \pbr\bfA(x^0),\Theta_{-,0}\bfE'(x^0)\pkt
    \right\}\right],
    \label{inn-inv-g}
    \eea
which is dynamically invariant. This fact may be checked directly
by differentiating the right-hand side of (\ref{inn-inv-g}) with
respect to $x^0$ and making use of (\ref{A-d}) or equivalently
(\ref{AE}).

By construction, (\ref{inn-inv-g}) gives the most general
positive-definite inner product that renders the generator of
time-translations as well as the helicity operator Hermitian. We
next impose the condition that (\ref{inn-inv-g}) be
Lorentz-invariant as well. This will give rise to the most general
physically admissible inner product on the space of Proca fields.

To quantify the requirement that the right-hand side of
(\ref{inn-inv-g}) be a Lorentz scalar we compute the inner product
of two plane-wave solutions of the Proca equation, i.e.,
(\ref{basic-sol}). Using (\ref{hel}), (\ref{k2-kdotS}),
(\ref{L-ep-pm}), (\ref{L-0-pm}), and (\ref{Theta-eph}) in
(\ref{inn-inv-g}), we find after a rather lengthy calculation
    \be
    \cbr A_{\ep,h}(\bfk),A_{\ep',h'}(\bfk')\para=
    \f{\kappa}{\cum}
    |N_{\ep,h}(\bfk)|^2\,\fa_{\ep,h}\:\omega_\bfk\,
    \delta^3(\bfk-\bfk')\delta_{\ep,\ep'}\delta_{h,h'},
    \label{basic-norm}
    \ee
where we have introduced the abbreviation
$\fa_{\ep,h}:=|\al_{\ep,h}|^2$. Because $\fa_{\ep,h}$ are positive
real numbers and the right-hand side of (\ref{basic-norm}) does
not involve $x^0$, this equation provides an explicit
demonstration of the invariance and positive-definiteness of the
inner product (\ref{inn-inv-g}).

Next, we recall from Refs.~\cite{weinberg,Yndurain} that
relativistically invariant normalization of two plane-wave
solutions is given by: $\cbr
A_{\ep,h}(\bfk;x),A_{\ep',h'}(\bfk';x)\ckt=2\,\omega_\bfk
\delta^3(\bfk-\bfk')\,\delta_{\ep,\ep'}\delta_{h,h'}$. This
together with the fact that $N_{\ep,h}(\bfk)$ must be a Lorentz
scalar so that $A_{\ep,h}(\bfk;x)$ is a four-vector, shows that
$\fa_{\ep,h}$ are \emph{dimensionless} positive real numbers,
i.e., they do not depend on $\bfk$ and obey the same Lorentz
transformation rule as scalars. This result ensures the
relativistic invariance of the inner product of any two solutions
$A$ and $A'$ having the general form:
    \be
    A=\sum_{\ep=\pm}\sum_{h=0,\pm1}\int_{\R^3} d^3\bfk\,c_{\ep,h}(\bfk)
    A_{\ep,h}(\bfk),
    \label{gen-form-sol}
    \ee
where $c_{\ep,h}(\bfk)$ are complex coefficients. In view of
(\ref{basic-norm}) and (\ref{gen-form-sol}) and the fact that the
inner product (\ref{inn-inv-g}) is a Hermitian sesquilinear form,
we obtain
    \be
    \cbr A,A'\para=\f{\kappa}{\cum}\sum_{\ep=\pm}\sum_{h=0,\pm1}
    \int_{\R^3} d^3\bfk\,\fa_{\ep,h}\,\omega_\bfk\,
    |N_{\ep,h}(\bfk)|^2\,c^*_{\ep,h}(\bfk)\,c'_{\ep,h}(\bfk).
    \label{inn-two-gen}
    \ee
Because $A$ and $A_{\ep,h}(\bfk)$ are four-vectors and
$d^3\bfk/\omega_\bfk$ is a relativistically invariant measure
\cite{weinberg,Yndurain}, $c_{\ep,h}(\bfk)$ obeys the same Lorentz
transformation rule as $\omega_\bfk^{-1}$. This in turn implies
that in order for the inner product (\ref{inn-two-gen}) to be
scalars, $\fa_{\ep,h}$ must transform as scalars. Therefore, in
view of (\ref{L-ep-pm}) and (\ref{L-0-pm}), we conclude that
$L^\pm_\pm$ and $L^0_\pm$ are dimensionless real numbers given by
    \be
    L^\ep_\pm=\f{1}{4}\left[\fa_{+,1}+\ep\fa_{+,-1}\pm
    \fa_{-,1}\pm\ep\fa_{-,-1}\right],~~~~~~~
    L^0_\pm=\f{1}{2}\left[\fa_{+,0}\pm\fa_{-,0}\right].
    \ee
Inserting these relations in (\ref{Theta-eph}) and using
(\ref{AE}) we find
    \[ \Theta_{+,+1}\bfA=\Theta_{+,0} D^{-1/2}\dot\bfE,~~~~~~~~
       \Theta_{+,-1}\bfE=-\Theta_{+,0} D^{-1/2}\dot\bfA.\]
These in turn allow us to express the most general physically
admissible inner product on the space $\cV$ of the Proca fields,
namely (\ref{inn-inv-g}), as
    \bea
    \cbr A,A'\para\equdef\f{\kappa}{2\cum}
    \left[\pbr\bfA(x^0),\Theta_{+,0}\,D^{-1/2}\dot\bfE'(x^0)\pkt-
    \pbr\bfE(x^0),\Theta_{+,0}\,D^{-1/2}\dot\bfA'(x^0)\pkt
    \right.\nn\\
    &&\vspace{2cm}\hspace{2.75cm}\left.-i\left\{
    \pbr\bfA(x^0),\Theta_{-,0}\bfE'(x^0)\pkt-
    \pbr\bfE(x^0),\Theta_{-,0}\bfA'(x^0)\pkt\right\}\right],
    \label{inn-inv-g-f}
    \eea
where, according to Eq.~(\ref{Theta-eph}),
    \be
    \Theta_{\pm,0}=[L^+_\pm-L^0_\pm]\fh^2+L^-_\pm\fh+L^0_\pm.
    \label{theta-pm0}
    \ee
Note that $L^+_+$ is a positive number and we can absorb it in the
definition of the field. Therefore, the inner product
(\ref{inn-inv-g-f}) actually involves five nontrivial free
parameters. Endowing $\cV$ with this inner product and Cauchy
completing the resulting inner product space we obtain a separable
Hilbert space which we denote by $\cHa$.

We can express the inner product (\ref{inn-inv-g-f}) in terms of
the indefinite Proca inner product (\ref{s3-inn}). First, we use
the unitary operator $U_{x^0_0}:\cHa\to\tcK$ to find an expression
for the action of the helicity operator
$\fH:=U_{x^0_0}^{-1}\Lambda U_{x^0_0}$ on a Proca field
$A\in\cHa$. $[\fH A](x^0)$ is a Proca field whose components can
be evaluated using (\ref{U-zero}), (\ref{div-grad}), and
(\ref{g-A-sol}). This gives
    \be
    [\fH A](x^0)=\left(0,\fh\bfA(x^0)\right).
    \label{helicity-A-exp}
    \ee
Because $[\Lambda,H]=0$  and $\fH$ and $\rh$ are respectively
related via a similarity transformation to $\Lambda$ and $H$, we
have $[\fH,\rh]=0$. This together with the identity $\fH^3=\fH$
suggest to use $\fH$ to split $\cV$ into the subspaces
$\cV_{\pm1}$ and $\cV_0$ of $\pm1$- and $0$-helicity Proca fields
according to $\cV_{h}:=\{A_{h}\in\cV | \fH
A_{h}=\mbox{\scriptsize$h$} A_{h}\}$ where
$\mbox{\scriptsize$h$}\in\{-1,0,+1\}$. We can use $\fH$ to define
the $\pm1$- and $0$-helicity components of the Proca fields
$A\in\cV$ according to:
    \be
    A_{\pm1}:=\frac{1}{2}\,(\fH^2 A \pm \fH A)\in \cV_{\pm1},
    ~~~~~~~A_0:=(A-\fH^2 A)\in \cV_0.
    \label{A-hel-dec}
    \ee
Clearly, $A=A_{+1}+A_{-1}+A_0$ which implies
$\cV=\cV_{+1}\oplus\cV_{-1}\oplus\cV_0$. Restricting the inner
product (\ref{inn-inv-g-f}) to $\cV_{\pm1}$ and $\cV_0$ (and
Cauchy completing the resulting inner product spaces) we obtain
Hilbert subspaces $\cH_{\{\fa\}\pm1}$ and $\cH_{\{\fa\}0}$ of
$\cHa$.

Next we recall that $\fH$ is obtained via a unitary similarity
transformation from the Hermitian operator $\Lambda$. This in turn
implies
    \be
    \cbr A,\fH A'\para=\cbr\fH A,A'\para.
    \label{hel-hermitian}
    \ee
Therefore, $\fH$ is a Hermitian operator acting is $\cHa$. It is a
physical observable that measures the helicity of $A$.
Furthermore, in light of (\ref{hel-hermitian}) and
(\ref{A-hel-dec}),
    \[\cbr A_{+1},A_{-1}\para=\cbr A_{+1},A_0\para=
    \cbr A_{-1},A_0\para=0,~~~~~~~~~~~
    \forall A_h\in\cV_h,~\mbox{\scriptsize$h$}\in\{-1,0,+1\}.\]
These relations show that
$\cHa=\cH_{\{\fa\}+1}\oplus\cH_{\{\fa\}-1}\oplus\cH_{\{\fa\}0}$.
Also, as we have shown in Section~\ref{CPT-section}, the chirality
operator ${\rm C}$ is a Hermitian involution acting in $\cHa$
which decomposes the Hilbert space $\cHa$ into its $\pm$-energy
Hilbert subspaces $\cH_{\{\fa\}\pm}$. Thus, using the operators
$\fH$ and ${\rm C}$, we can decompose the Hilbert space into six
mutually orthogonal subspaces $\cH_{\{\fa\}(\ep,h)}$, where
$\ep=\{+,-\}$, and $\mbox{\scriptsize$h$}\in\{-1,0,+1\}$.
Employing this decomposition of $A$ in (\ref{inn-inv-g-f}), we
find
    \be
    \cbr A,A'\para=\sum_{\ep=\pm}\sum_{h=0,\pm1}\ep\,\fa_{\ep,h}\,
    \cbr A_{\ep,h},A'_{\ep,h}\ckt_{_{\Sigma_3}}.
    \label{inn-g-of-s3}
    \ee
Therefore, (\ref{inn-inv-g-f}) is a linear combination of the
indefinite inner products (\ref{s3-inn}) of the components
$A_{\ep,h}$ of $A$ with suitable non-negative coefficients. Note
that the decomposition of the fields into the helicity components
(\ref{A-hel-dec}) is clearly frame-dependent and so
(\ref{inn-g-of-s3}) hides the Lorentz covariance of the theory.
But the inner product (\ref{inn-inv-g-f}) does not involve the
explicit splitting of the fields into $(\ep,h)$-components.

In the remainder of this section, we obtain the Foldy
transformation associated with the general inner product
(\ref{inn-inv-g-f}). First, consider the invertible linear
operator $\cA$ of (\ref{cA-g}). This is a unitary operator mapping
$\tcK$ onto $\cK$, for it satisfies, for all $\xi,\zeta\in\tcK$,
    \be
    \bbr\cA\xi,\cA\zeta\kkt_{\eta_+}=\br\cA\xi,\eta_+\cA\zeta\kt
    =\br\xi,\cA^\dagger\eta_+\cA\zeta\kt=
    \bbr\xi,\zeta\kkt_{\teta_+}.
    \ee
We can find the explicit form of $\cA$ and $\cA^{-1}$ by
substituting (\ref{H-evec}) and (\ref{Hdag-evec}) in (\ref{cA-g}).
This gives
    \bea
    \cA\equa[F^+_-\sigma_3M-F^0_-\sigma_3N+(F^+_+-F^0_+)\sigma_0]\fh^2
    +[F^-_-\sigma_3M+F^-_+\sigma_0]\fh+F^0_-\sigma_3N+F^0_+\sigma_0,~~~
    \label{cA}\\
    \cA^{-1}\equa[\tF^+_-\sigma_3M-\tF^0_-\sigma_3N+
    (\tF^+_+-\tF^0_+)\sigma_0]\fh^2
    +[\tF^-_-\sigma_3M+\tF^-_+\sigma_0]\fh+
    \tF^0_-\sigma_3N+\tF^0_+\sigma_0,~~~
    \label{cA-inv}
    \eea
where
    \bea
    F^\ep_\pm\equdef\f{1}{2}\left[Z^+_\ep\pm Z^-_\ep\right],~~~~~
    F^0_\pm:=\f{1}{2}\left[\al_{+,0}\pm\al_{-,0}\right],\\
    \tF^\ep_\pm\equdef\f{1}{2}\left[\tZ^+_\ep\pm \tZ^-_\ep\right],~~~~~
    \tF^0_\pm:=\f{1}{2}\left[\al_{+,0}^{-1}\pm\al_{-,0}^{-1}\right],\\
    Z^\ep_\pm&:=&\f{1}{2}[\al_{\ep,1}\pm\al_{\ep,-1}],~~~~~
    \tilde{Z}^\ep_\pm:=\f{1}{2}[\al^{-1}_{\ep,1}\pm\al^{-1}_{\ep,-1}].
    \eea
Next, we define $\trho:\tcK\to\cH'$ and $\cUa:\cHa\to\cH'$ by
    \bea
    \trho&:=&\rho\,\cA,
    \label{trho}\\
    \cUa&=&\trho\, U_{x^0_0}.
    \label{cUa}
    \eea
In view of $\rho^\dagger=\rho=\sqrt\eta_+$ and (\ref{eta-+g}),
$\trho$ is a unitary transformation mapping $\tcK$ onto $\cH'$,
    \be
    \br\trho\xi,\trho\zeta\kt=\br\xi,\trho^\dagger\trho\zeta\kt
    =\br\xi,\cA^\dagger\eta_+\cA\zeta\kt=\bbr\xi,\zeta\kkt_{\teta_+}.
    \ee
This in turn implies that $\cUa:\cHa\to\cH'$ is also a unitary
transformation,
    \be
    \br\cUa A,\cUa A'\kt=\cbr A,A'\para,~~~~~~~~~~
    \forall A, A'\in\cHa.
    \label{unitary-u-gen}
    \ee

We can compute the explicit form of the unitary operator $\cUa$
and its inverse. This requires computing $\trho$ and its inverse.
Substituting (\ref{rho}), (\ref{rho-ij}), (\ref{cA}) and
(\ref{cA-inv}) in (\ref{trho}), we have
    \be
    \trho=\f{1}{2\cum\sqrt{\ld}}\,\left(\begin{array}{cc}
    \trho_{+,+} & \trho_{-,+} \\
    \trho_{-,-} & \trho_{+,-} \\
    \end{array}\right),~~~~~~
    \trho^{-1}=\f{1}{2\cum\sqrt{\ld}}\,\left(\begin{array}{cc}
    \trho_{+,+}^{~{\rm inv}} & -\trho_{-,-}^{~{\rm inv}} \\
    -\trho_{-,+}^{~{\rm inv}} & \trho_{+,-}^{~{\rm inv}} \\
    \end{array}\right),
    \label{trho-trhoinv}
    \ee
where the entries are the following operators that act in $\tcH$.
    \bea
    \tilde{\rho}_{\ep,\ep'}\equdef
    \cum\,(\ld D^{1/4}+\ep D^{-1/4})
    \left[Z^{\ep'}_+\fh^2+Z^{\ep'}_-\fh\right]+
    \al_{\ep',0}\,(\ld\cum^2D^{-1/4}+\ep D^{1/4})[\lnd_0-\fh^2],
    \label{trho-ij}\\
    \trho_{\ep,\ep'}^{~{\rm inv}}\equdef
    \cum\,(\ld D^{1/4}+\ep D^{-1/4})
    \left[\tilde{Z}^{\ep'}_+\fh^2+\tilde{Z}^{\ep'}_-\fh\right]+
    \al_{\ep',0}^{-1}\,(\ld\cum^2D^{-1/4}+\ep D^{1/4})[\lnd_0-\fh^2].
    \label{trho-inv-ij}
    \eea
Next, we use (\ref{cUa}), (\ref{trho-trhoinv}), (\ref{trho-ij}),
(\ref{U-zero}), and (\ref{6-comp-psi}), to find, for all
$A\in\cH$,
    \be
    \cUa A=\frac{1}{2}\sqrt{\f{\kappa}{\cum}}\,\left(\begin{array}{c}
    \fU_{+,+}\,\bfA(x^0_0)-i\,\fU_{+,-}\,\bfE(x^0_0)\\
    \fU_{-,+}\,\bfA(x^0_0)+i\,\fU_{-,-}\,\bfE(x^0_0)
    \end{array}\right),
    \label{tu-A=}
    \ee
where the operators $\fU_{\ep,\ep'}:\tcH\to\tcH$, are given by
    \be
    \fU_{\ep,\ep'}=\left[Z^\ep_+ D^{\ep'/4}-\cum^{\ep'}
    \al_{\ep,0} D^{-\ep'/4}\right] \fh^2+
    Z^\ep_- D^{\ep'/4} \fh+\cum^{\ep'} \al_{\ep,0} D^{-\ep'/4}.
    \ee

Again the arbitrary parameter $\ld$ drops out of the calculations.
$\cUa^{-1}=U_{x^0_0}^{-1}\tilde\rho^{-1}$ is also easy to
calculate. Let $\xi\in{\cal H}'$ be a six-component vector (as in
(\ref{xi-zeta})) with components $\bfxi^1$ and $\bfxi^2$. Then in
view of (\ref{tu-A=}), $\cUa^{-1}\xi$ is the Proca field
$A\in\cHa$ satisfying the following initial conditions:
    \bea
    A^0(x^0_0)\equa \,\f{D^{-1/4}}{\sqrt{\cum\kappa}}\,
    \bffK\cdot\left(\f{\bfxi^1}{\al_{+,0}}-
    \f{\bfxi^2}{\al_{-,0}}\right),
    ~~~~~
    \dot A^0(x^0_0)=-i\,\f{D^{1/4}}{\sqrt{\cum\kappa}}\,
    \bffK\cdot\left(\f{\bfxi^1}{\al_{+,0}}+
    \f{\bfxi^2}{\al_{-,0}}\right),
    \label{tU-inverse-0} \\
    \bfA(x^0_0)\equa\sqrt{\f{\cum}{\kappa}}\,
    \left(\fU_{+,+}^{-1}\bfxi^1+\fU_{-,+}^{-1}\bfxi^2\right),
    ~~~~~~
    \dot\bfA(x^0_0)=-i\,\sqrt{\f{\cum}{\kappa}}\,
    D^{1/2}\,\left(\fU_{+,+}^{-1}\bfxi^1-
    \fU_{-,+}^{-1}\bfxi^2\right),
    \label{tU-inverse-i}
    \eea
where $\fU_{\ep,+}^{-1}$ is the inverse of $\fU_{\ep,+}$ given by
    \be
    \fU_{\ep,+}^{-1}=\left[\tZ^\ep_+ D^{-1/4}-
    \f{D^{1/4}}{\cum\al_{\ep,0}}\right] \fh^2+
    \tZ^\ep_- D^{-1/4} \fh+\f{D^{1/4}}{\cum\al_{\ep,0}},
    \ee
and we have made use of (\ref{constraints}), (\ref{AE}),
(\ref{div-grad}), (\ref{6-comp-psi}), (\ref{U-zero}), and
(\ref{trho-trhoinv}) -- (\ref{trho-inv-ij}). By virtue of
(\ref{g-A-sol}), for all $x^0_0\in\R$, we have
    \bea
    A^0(x^0)&=&\,\f{D^{-1/4}}{\sqrt{\cum\kappa}}\,
    \left[e^{-i(x^0-x^0_0) D^{1/2}}\bffK\cdot\f{\bfxi^1}{\al_{+,0}}-
    e^{i(x^0-x^0_0) D^{1/2}}\bffK\cdot\f{\bfxi^2}{\al_{-,0}}\right],\\
    \bfA(x^0)&=&\sqrt{\f{\cum}{\kappa}}\,
    \left[e^{-i(x^0-x^0_0) D^{1/2}}\fU_{+,+}^{-1}\bfxi^1+
    e^{i(x^0-x^0_0) D^{1/2}}\fU_{-,+}^{-1}\bfxi^2\right].
    \eea

Next we recall that since $[H,\cA]=0$, we have
$\rh_{\{\fa\}}=\rh$. Therefore, in light of the above analysis,
the pairs $(\cHa,\rh)$, $(\tcK,H)$, and $(\cH',H')$ are unitarily
equivalent; they represent the same quantum system.

Finally, we give an alternative form of the Foldy transformation
(\ref{tu-A=}). In view of (\ref{A-pm-field}) and
$\fU_{\ep,-}^{-1}\bfE = -\,\fU_{\ep,+}\,D^{-1/2}\dot\bfA$, we have
    \be
    \cUa A=\sqrt{\f{\kappa}{\cum}}\,\left(\begin{array}{c}
    \fU_{+,+} \bfA_+(x^0_0)\\
    \fU_{-,+} \bfA_-(x^0_0)
    \end{array}\right).
    \label{tu-A=2}
    \ee
In particular, setting $\al_{\ep,h}=1$, for all $\ep\in\{+,-\}$
and $\mbox{\scriptsize$h$}\in\{-1,0,+1\}$, yields
    \be
    \cU A=\sqrt{\f{\kappa}{\cum}}\,\fU\left(\begin{array}{c}
    \bfA_+(x^0_0)\\
    \bfA_-(x^0_0)
    \end{array}\right),
    \label{u-A=2}
    \ee
where $\bfA_\pm$ is $\pm$-energy component of Proca field. These
equations show that $\fU_{\ep,+} \bfA_\ep$ (specially
$\fU\bfA_\ep$) satisfy the Foldy equation \cite{Foldy,Case}
    \be
    i\p_0\fU_{\ep,+} \bfA_\ep=\ep D^{1/2}
    \fU_{\ep,+} \bfA_\ep.
    \label{foldy}
    \ee

\section{Conserved Current Density} \label{cons-current-sec}

The relativistic and time-translation invariance of the
positive-definite inner products (\ref{inn-inv-g-f}) suggest the
existence of an associated conserved four-vector current density
$J\suba^\mu$. In this section we use the approach of
\cite{MZ-ann06-1} to compute $J\suba^\mu$ for the case
$\fa_{\ep,h}=1$ and obtain a manifestly covariant expression for
$J\suba^\mu$ with $\fa_{\ep,h}=1$ which for brevity we denote by
$J^\mu$.

According to (\ref{inn-inv-g-f}), for all $A\in\cV$,
    {\small\bea
    \cbr A,A\para\equa\frac{\kappa}{2\cum}\!\int_{\R^3}\!\!\!d^3\bfx
    \left\{\bfA(x^0,\bfx)^*\cdot\!\left(\Theta_{+,0}D^{-1/2}\dot\bfE(x^0,\bfx)\right)
    -\bfE(x^0,\bfx)^*\cdot\!\left(\Theta_{+,0}D^{-1/2}\dot\bfA(x^0,\bfx)\right)\right.\nn\\
    &&\vspace{3cm}\hspace{2cm}\left.-i\left[
    \bfA(x^0,\bfx)^*\cdot\!\left(\Theta_{-,0}\bfE(x^0,\bfx)\right)-
    \bfE(x^0,\bfx)^*\cdot\!\left(\Theta_{-,0}\bfA(x^0,\bfx)\right)
    \right]\right\}.
    \label{i-exp}
    \eea}
In analogy with non-relativistic QM, we define the $0$-component
of the current density $J\suba^\mu$ associated with $A$ as the
integrand in (\ref{i-exp}), i.e.,
    \bea
    J\suba^0(x)\equdef\frac{\kappa}{2\cum}\left\{
    \bfA(x)^*\cdot\left(\Theta_{+,0}D^{-1/2}\dot\bfE(x)\right)
    -\bfE(x)^*\cdot\left(\Theta_{+,0}D^{-1/2}\dot\bfA(x)\right)\right.\nn\\
    &&\vspace{3cm}\hspace{2cm}\left.-i\left[
    \bfA(x)^*\cdot\left(\Theta_{-,0}\bfE(x)\right)-
    \bfE(x)^*\cdot\left(\Theta_{-,0}\bfA(x)\right)
    \right]\right\}.
    \label{j-zero-gen}
    \eea
Setting $\fa_{\ep,h}=1$ for all $\ep\in\{+,-\}$ and
$\sh\in\{-1,0,+1\}$, yields
     \be
    J^0(x):=\frac{\kappa}{2\cum}\left\{
    \bfA(x)^*\cdot D^{-1/2}\dot\bfE(x)
    -\bfE(x)^*\cdot D^{-1/2}\dot\bfA(x)\right\}.
    \label{j-zero}
    \ee
In order to obtain the spatial components of $J^\mu$, we follow
the procedure outlined in Ref.~\cite{MZ-ann06-1}. Namely, we
perform an infinitesimal Lorentz boost transformation that changes
the reference frame to the one moving with a velocity
$\mathbf{v}$:
    \be
    x^0\to {x'}^0=x^0-\bfbeta\cdot\bfx,~~~~~~~~~~
    \bfx\to\bfx'=\bfx-\bfbeta x^0,
    \label{boost}
    \ee
where $\bfbeta:=\mathbf{v}/c$. Assuming that $J^\mu$ is indeed a
four-vector field and neglecting the second and higher order terms
in powers of the components of $\bfbeta$, we then find
    \be
    J^0(x)\to{J'}^0(x')=J^0(x)-\bfbeta\cdot\bfJ(x).
    \label{boost-j}
    \ee
Next, we use (\ref{j-zero}) to obtain
    \be
    {J'}^0(x'):=\frac{\kappa}{2\cum}\left\{
    \bfA'(x')^*\cdot D'^{-1/2}\dot\bfE'(x')
    -\bfE'(x')^*\cdot D'^{-1/2}\dot\bfA'(x')\right\},
    \label{j-zero-prime}
    \ee
where $x':=({x'}^0,\bfx')$, $D'=-{\nabla'}^2+\cum^2$,
$\bfE'(x')=-\dot\bfA'(x')-\bfdel' {A'}^0(x')$, and
$\dot\bfA'(x'):=\p\bfA'(x')/\p{x'}^0$. This reduces the
determination of $\bfJ$ to expressing the right-hand side of
(\ref{j-zero-prime}) in terms of the quantities associated with
the original (unprimed) frame and comparing the resulting
expression with (\ref{boost-j}).

Under the transformation (\ref{boost}), the four-vector
$A=(A^0,\bfA)$ transforms as
    \be
    A^0(x)\to {A'}^0(x')=A^0(x)-\bfbeta\cdot\bfA(x),~~~~~~~~~~
    \bfA(x)\to\bfA'(x')=\bfA(x)-\bfbeta A^0(x).
    \label{boost-A}
    \ee
Therefore, in view of (\ref{boost}), we can easily obtain the
transformation rule for $\bfE$. The result is
    \be
    \bfE(x)\to\bfE'(x')=\bfE(x)+\bfbeta\times(\bfdel\times\bfA(x)).
    \label{boost-E}
    \ee
Because the chirality operator ${\rm C}=i D^{-1/2}\p_0$ is Lorentz
invariant \cite{MZ-ann06-1},
    \be
    D'^{-1/2}\p'_0= D^{-1/2}\p_0.
    \label{boost-C}
    \ee
Substituting (\ref{boost-A}) -- (\ref{boost-C}) in
(\ref{j-zero-prime}), making use of (\ref{boost-j}), we obtain
    \bea
    J^i(x)\equa\frac{\kappa}{2\cum}\left\{
    A^*(x)\cdot\p^i D^{-1/2}\dot A(x)-
    [\p^i A(x)^*]\cdot D^{-1/2}\dot A(x)-\right.\nn\\
    & &\vspace{2cm}\hspace{1.2cm}\left.[A^*(x)\cdot\p] D^{-1/2}\dot A^i(x)+
    [ D^{-1/2}\dot A(x)\cdot\p] A^i(x)^*\right\},
    \label{j}
    \eea
where $\p:=(\p^0,\bfdel)$ and for any two four-vectors $v_1$ and
$v_2$, $v_1\cdot v_2:=v_1^\mu v_{2\mu}$. This relation suggests
    \bea
    J^\mu(x)\equa\frac{\kappa}{2\cum}\left\{
    A^*(x)\cdot\p^\mu D^{-1/2}\dot A(x)-
    [\p^\mu A(x)^*]\cdot D^{-1/2}\dot A(x)-\right.\nn\\
    & &\vspace{2cm}\hspace{1.2cm}\left.[A^*(x)\cdot\p] D^{-1/2}\dot A^\mu(x)+
    [ D^{-1/2}\dot A(x)\cdot\p] A^\mu(x)^*\right\}.
    \label{J}
    \eea
It is not difficult to check (using the Proca equation) that the
expression for $J^0$ obtained using this equation agrees with the
one given in (\ref{j-zero}).

We can use (\ref{F-mu-nu}) and (\ref{A-c}) to further simplify
(\ref{J}). This yields
    \bea
    J^\mu(x)\equa\frac{i\kappa}{2\cum}\left\{
    {A}_\nu(x)^* F_c^{\nu\mu}(x) - {F}^{\nu\mu}(x)^*
    A_{c\,\nu}(x)\right\},
    \label{J-2}
    \eea
where $F_c^{\mu\nu}:=\p^\mu A_c^\nu-\p^\nu A_c^\mu$. The current
density $J^\mu$ which is generally complex-valued has the
following remarkable properties:
    \begin{enumerate}
    \item The expression~(\ref{J-2}) for $J^\mu$ is manifestly
    covariant; since $A$ and $A_c$ are four-vector fields, so is
    $J^\mu$.
    \item Using the fact that both $A$ and $A_c$
    satisfy the Proca equation (\ref{A-proca}), one easily checks
    that $J^\mu$ satisfies the following continuity equation.
        \be
        \p_\mu J^\mu=0.
        \label{conti-2}
        \ee
    Hence it is a conserved current density.
    \end{enumerate}

Next, we use (\ref{J-2}) to derive a manifestly covariant
expression for the inner product (\ref{inn-inv-2}) on the space of
solutions of the Proca equation~(\ref{A-d}). The result is
    \bea
    \cbr A,A'\ckt\equa\frac{i\kappa}{2\cum}
    \int_{\sigma} d\sigma(x)~ n_\mu(x)\left\{
    {A}_\nu(x)^* F_c'^{\nu\mu}(x) -{F}^{\nu\mu}(x)^*
    A'_{c\,\nu}(x)\right\},
    \label{man-cov}
    \eea
where $\sigma$ is an arbitrary spacelike (Cauchy) hypersurface of
the Minkowski space with volume element $d\sigma$ and unit
(future) timelike normal four-vector $n^\mu$. Note that in
deriving~(\ref{man-cov}) we have also made an implicit use of the
fact that any inner product is uniquely determined by the
corresponding norm \cite{kato}.

Using the same approach we have calculated $J\suba^\mu$ for
$\fa_{\ep,h}\neq 1$ and checked that indeed it is a conserved
complex-valued four-vector field. But we were not able to obtain a
manifestly covariant expression for $J\suba^\mu$ in this case. As
the expression for $J\suba^\mu$ is highly complicated we do not
present it here.

\section{Physical Observables and Wave Functions for Proca
Fields}\label{phy-obs-sec}

The unitary equivalence of the representations $(\cHa,\rh)$,
$(\tcK,H)$, and $(\cH',H')$ for the quantum system describing the
Proca fields allows for the construction of the observables of
this system using any of these representations. Because ${\cal
H}'=\tcH\oplus \tcH$ and $\tcH=L^2(\R^3)\oplus L^2(\R^3)\oplus
L^2(\R^3)$, the Foldy representation $({\cal H}',H')$ is more
convenient for this purpose. In this section we construct the
observables in this representation and use the unitary map
$\cUa:\cHa\to\cH'$ to obtain their form in the standard
(covariant) representation $(\cHa,\rh)$. Again we follow closely
the approach used in \cite{M-IJMPA} to construct the observables
for KG fields.

In the Foldy representation, we introduce the following set of
basic observables
    \be
    \bfX'_\fm:=\bfrmx\otimes\Sigma_\fm,~~~~~~
    \bfP'_\fm:=\bfrmp\otimes\Sigma_\fm,~~~~~~
    \cS'_\fm:=1\otimes\Sigma_\fm,
    \label{ob-0}
    \ee
where, $\bfrmx$, $\bfrmp=\hbar\,\bffK$, and $1$ are respectively
the position, momentum, and identity operators acting in
$L^2(\R^3)$, $\fm\in\{0, 1, 2, \cdots, 35\}$, and $\Sigma_\fm$'s
are given in Eq.~(\ref{Sigma-basis}). In the following, we will
omit `$1\otimes$' for brevity. In particular, we will identify
$\cS'_\fm$ with $\Sigma_\fm$.

As we mentioned in Section~1, the operators $\bfX'_0$, $\Sigma_3$,
$\Sigma_{12}$, $\Sigma_{15}$, $\Sigma_{32}$, and $\Sigma_{35}$
form a maximal commuting set of observables acting in $\cH'$. In
particular, we can use their common eigenvectors, namely
    \be
    \xi^{(\ep,s)}_\bfx:=|\bfx\kt\otimes e_{\ep,s},~~~~~~~
    \bfx\in\R^3,~~\epsilon\in\{-,+\},~~s\in\{-1,0,+1\},
    \label{xi}
    \ee
to construct a basis of $\cH'$. In (\ref{xi}), $e_{\ep,s}$ are the
vectors defined in (\ref{com-eg-ve-Z}) and $|\bfx\kt$ are the
$\delta$-function normalized position kets satisfying $
\bfrmx|\bfx\kt=\bfx|\bfx\kt$,
$\br\bfx|\bfx'\kt=\delta^3(\bfx-\bfx')$, and
$\int_{\R^3}d^3\bfx\:|\bfx\kt\br\bfx|=1$. It is easy to see that
indeed $\bfX'_0\,\xi^{(\ep,s)}_\bfx\!=\bfx \,\xi^{(\ep,s)}_\bfx$,
$\Sigma_3\,\xi^{(\ep,s)}_\bfx \!=\ep\,\xi^{(\ep,s)}_\bfx$, and
$\Sigma_{12}\,\xi^{(\ep,s)}_\bfx \!=s\,\xi^{(\ep,s)}_\bfx$.
Furthermore,
    \be
    \br \xi^{(\ep,s)}_\bfx,\xi^{(\ep',s')}_{\bfx'}\kt=
    \delta_{\ep,\ep'}\delta_{s,s'}\delta^3(\bfx-\bfx'),~~~~~~~~~
    \sum_{\ep=\pm}\sum_{s=0,\pm1}\int_{\R^3}d^3\bfx
    \:|\xi^{(\ep,s)}_\bfx\kt\br\xi^{(\ep,s)}_\bfx|=\Sigma_0.
    \label{orthonormal}
    \ee

We can express any six-component vector $\Psi'\in{\cal H}'$ in the
basis $\{\xi^{(\ep,s)}_\bfx\}$ according to
    \be
    \Psi'=\sum_{\ep=\pm}\sum_{s=0,\pm1}\int_{\R^3}d^3\bfx\:
    f(\ep,s,\bfx)\,\xi^{(\ep,s)}_\bfx,
    \label{wf-1}
    \ee
where $f:\{-,+\}\times\{-1,0,+1\}\times\R^3\to\C$ is the wave
function associated with $\Psi'$ in the position-representation,
i.e.,
    \be
    f(\ep,s,\bfx):=\br\xi^{(\ep,s)}_\bfx,\Psi'\kt.
    \label{wf-0}
    \ee
Also, the action of a physical observable $O':\cH'\to\cH'$ on the
state vector $\Psi'\in\cH'$ is given by
    \be
    O'\Psi'=\sum_{\ep=\pm}\sum_{s=0,\pm1}\int_{\R^3}d^3\bfx\:
    [\hat O f(\ep,s,\bfx)]\,\xi^{(\ep,s)}_\bfx,
    \label{ob-1}
    \ee
where
    \be
    \hat O f(\ep,s,\bfx):=
    \sum_{\ep'=\pm}\sum_{s'=0,\pm1}\int_{\R^3}d^3\bfx'
    f(\ep',s',\bfx')\,\br\xi^{(\ep,s)}_\bfx,O'\xi^{(\ep',s')}_{\bfx'}\kt.
    \label{ob-2}
    \ee
This provides the representation of observables in terms of linear
operators acting on the square-integrable wave functions $f$.

Next, we introduce the operators
    \be
    \xafm:=\cUa^{-1}\,\bfX'_\fm\,\cUa,~~~~~~~
    \pafm:=\cUa^{-1}\,\bfP'_\fm\,\cUa,~~~~~~~
    \safm:=\cUa^{-1}\,\Sigma_\fm\,\cUa,
    \label{ob-4}
    \ee
that act in $\cHa$, and define the Proca fields
    \be
    \Aax^{(\ep,s)}=( \Aax^{0(\ep,s)},\bfAax^{(\ep,s)})
    :=\cUa^{-1}\,\xi^{(\ep,s)}_\bfx,
    \label{local}
    \ee
which form a complete orthonormal basis of $\cHa$:
    \be
    \cbr \Aax^{(\ep,s)},\Aaxp^{(\ep',s')}\para=
    \delta_{\ep,\ep'} \delta_{s,s'} \delta^3(\bfx-\bfx'),~~~~~~~
    \sum_{\ep=\pm} \sum_{s=0,\pm1} \int_{\R^3}d^3\bfx
    \:|\Aax^{(\ep,s)})(\Aax^{(\ep,s)}|=s_{0\{\fa\}}.
    \label{orthonormal-A}
    \ee
Here we have used (\ref{unitary-u-gen}), (\ref{orthonormal}), and
(\ref{local}), $s_{0\{\fa\}}$ coincides with the identity operator
for $\cHa$, and for all $A,A'\in\cHa$, the operator $|A)(A'|$ is
defined by $|A)(A'|A'':=\cbr A',A''\para A$, for any $A''\in\cHa$.
In view of (\ref{orthonormal-A}) and
    \be
    \cbr \Aax^{(\ep,s)},A\para=\br\cUa \Aax^{(\ep,s)},
    \cUa A\kt=\br\xi^{(\ep,s)}_\bfx,\Psi'\kt=f(\ep,s,\bfx),
    \label{f}
    \ee
we can express any Proca field $A\in\cHa$ in the basis
$\{\Aax^{(\ep,s)}\}$ according to
    \be
    A=\sum_{\ep=\pm} \sum_{s=0,\pm1} \int_{\R^3}d^3\bfx\:
    f(\ep,s,\bfx)\,\Aax^{(\ep,s)}.
    \label{wf}
    \ee
Here we do not label the wave functions $f$ with the subscript
$\{\fa\}$, because they do not depend on the choice of the
parameter $\fa_{\ep,h}$. The proof of this assertion uses the
unitary operator
    \be
    \Ua:=\cUa^{-1}\cU,
    \label{cur-u-a}
    \ee
that maps $\cH$ onto $\cHa$ and is identical with the one given in
\cite{MZ-ann06-1} for the KG fields.

The physical observables $o_{\{\fa\}}:\cHa\to\cHa$ are uniquely
specified in terms of their representation in the basis
$\{\Aax^{(\ep,s)}\}$; for all $A\in\cHa$
    \be
    o_{\{\fa\}}A=\sum_{\ep=\pm} \sum_{s=0,\pm1} \int_{\R^3}d^3\bfx
    [\hat O f(\ep,s,\bfx)]\Aax^{(\ep,s)},
    \label{ob-5}
    \ee
where $\hat O f(\ep,s,\bfx)$ is defined by (\ref{ob-2}). This
follows from
$\cbr\Aax^{(\ep,s)},o_{\{\fa\}}\Aaxp^{(\ep',s')}\para=
\br\xi^{(\ep,s)}_\bfx,O'\xi^{(\ep',s')}_{\bfx'}\kt$. In view of
(\ref{orthonormal-A}) and (\ref{wf}), we can express the
transition amplitudes between two states (the inner product of two
Proca fields) in the form
    \be
    \cbr A,A'\para=\sum_{\ep=\pm}\sum_{s=0,\pm1}\int_{\R^3}d^3\bfx\:
        f(\ep,s,\bfx)^* f'(\ep,s,\bfx).
    \label{inn-of-f}
    \ee
More generally, for any observable $o_{\{\fa\}}:\cHa\to\cHa$, we
have
    \bea
    \cbr A,o_{\{\fa\}}A'\para\equa\sum_{\ep=\pm}\sum_{s=0,\pm1}\int_{\R^3}d^3\bfx\:
        f(\ep,s,\bfx)^*\, \hat O f'(\ep,s,\bfx)\nn\\
        \equa\sum_{\ep=\pm}\sum_{s=0,\pm1}\int_{\R^3}d^3\bfx\:
        [\hat O f(\ep,s,\bfx)]^* f'(\ep,s,\bfx).
    \label{ob-7}
    \eea

The above discussion shows that we may view the wave functions $f$
as elements of ${\cal H}'$, and similarly to the case of KG fields
\cite{M-ann,M-IJMPA}, formulate the QM of Proca fields in terms of
these wave functions. In this formulation the observables are
Hermitian operators $\hat O$ acting on the wave functions. For
example, the action of $\bfrmx_{0\{\fa\}}$, $\bfrmp_{0\{\fa\}}$,
$s_{3\{\fa\}}$, and $s_{12\{\fa\}}$ on $A$ corresponds to the action
of the operators $\hat{\bfrmx}_{0\{\fa\}}$,
$\hat{\bfrmp}_{0\{\fa\}}$, $\hat s_{3\{\fa\}}$, and $\hat
s_{12\{\fa\}}$ on $f$, where
    \bea
    \hat{\bfrmx}_{0\{\fa\}} f(\ep,s,\bfx)&:=&\bfx f(\ep,s,\bfx),
    ~~~~~~~~~~
    \hat{\bfrmp}_{0\{\fa\}} f(\ep,s,\bfx):=-i\hbar\bfdel
    f(\ep,s,\bfx),\label{wf-ob1}\\
    \hat s_{3\{\fa\}} f(\ep,s,\bfx)&:=&\ep f(\ep,s,\bfx),
    ~~~~~~~~~~
    \hat s_{12\{\fa\}} f(\ep,s,\bfx):=s f(\ep,s,\bfx).
    \label{wf-ob2}
    \eea
These equations show that $f(\ep,s,\bfx)$ are the position wave
functions with definite chirality (sign of the energy) $\ep$ and
spin $s$ (say along the $x^3$-direction). They are however not the
eigenfunctions of the helicity operator. To see this, we recall
from (\ref{inn-p-6}) that $\cH'=\tcH\oplus\tcH$, and the spin
operator acting in $\tcH$ is given by $\bfS$ of Eq.~(\ref{S123}).
This shows that the spin operator acting in $\cH'$
is\footnote{Throughout this paper we express spin and angular
momentum operators in units of $\hbar$.}
    \be
    \bfS':=(\Sigma_{28},-\Sigma_{20},\Sigma_8)
    =\left(\begin{array}{cc}
    \bfS & 0 \\
    0 & \bfS \end{array}\right).
    \label{spin-foldy}
    \ee
Denoting the spin operator acting in $\cHa$ by $\bfs_{\{\fa\}}$,
we have
    \be
    \bfs_{\{\fa\}}=\cUa^{-1}\,\bfS'\,\cUa.
    \label{spin-gen-def}
    \ee
The projection of this operator along the momentum
$\bfrmp_{0\{\fa\}}$ gives the helicity operator:
    \be
    (\hat{\bfrmp}_{0\{\fa\}}\cdot\hat{\bfs}_{\{\fa\}}) f(\ep,s^i,\bfx)=
    \varepsilon_{ijk}\: \p_j\: f(\ep,s^k,\bfx),
    \label{f-helicity}
    \ee
where $(s^1,s^2,s^3)=(+1,-1,0)$, and we employed (\ref{ob-2}).

Similarly, the action of the Hamiltonian $\rh$ on $A$ corresponds to
the action of the operator $\hat \rh:=\hat
s_{3\{\fa\}}\sqrt{\hat{\bfrmp}_{0\{\fa\}}^2+m^2c^2}$ on the wave
function $f$:
    \be
    \hat{\rh}f(\ep,s,\bfx)=
    \hbar\ep\sqrt{-\nabla^2+\cum^2}f(\ep,s,\bfx).
    \label{h-hat=}
    \ee
Subsequently, in view of (\ref{h-hat=}), and (\ref{sch-eq-h}), the
dynamics of the evolving Proca field $A_{x^0}$ is determined in
terms of the wave functions $f(\ep,s,\bfx;x^0)=\cbr
\Aax^{(\ep,s)},A_{x^0}\para$, according to
    \be
    i\hbar\p_0 f(\ep,s,\bfx;x^0)=
    \ep\sqrt{-\hbar^2\nabla^2+m^2c^2}\; f(\ep,s,\bfx;x^0).
    \label{sch-eq-f}
    \ee
Furthermore, applying $i\p_0$ to both sides of (\ref{sch-eq-f}),
we can check that the wave functions $f$ also satisfies the KG
equation: $[\p_0^2-\nabla^2+\cum^2]f(\ep,s,\bfx;x^0)=0$.

Next, recall that because the time-reversal operator
(\ref{PT-prime}) acting in $\cH'$ commutes with $\bfX'_0$, the
eigenvectors $\xi^{(\ep,s)}_\bfx$ may be taken to be real. In this
case the action of the time-reversal operator $T={\rm PT}$ on any
$A\in\cHa$ is equivalent to the complex-conjugation of the
associated wave-function $f$, i.e., $\hat T
f(\ep,s,\bfx)=f(\ep,s,\bfx)^*$. Similarly, we can identify the
operators $\hat s_3$ and $\hat s_{12}$, respectively, with the
chirality and spin operators acting on the wave functions $f$.

As we shall see in the following section the wave functions
$f(\ep,s,\bfx)$ furnish a position representation for the QM of
the Proca fields. The corresponding position operator is the
spin-1 analog of the Newton-Wigner position operator for the KG
field \cite{ne-wi,localizations} and similarly to the latter fails
to be relativistically covariant.\footnote{See however
\cite{Farkas}.} This means that the above-mentioned
position-representation provides a non-covariant description of a
quantum system that also admits a unitary-equivalent covariant
description in terms of the Hilbert space $\cHa$ and the
Hamiltonian $\rh$.

\section{Position, Spin, and Localized States}

As discussed in \cite{M-IJMPA} for the KG fields, the canonical
quantization scheme that provides the physical meaning of the
observables yields the Foldy representation of the quantum
system.\footnote{In the classical limit, each component of a Proca
field corresponds to a classical free particle of energy
$E=\pm\sqrt{p^2+m^2c^2}$. Upon quantization
$E\to\pm\hbar\sqrt{-\nabla^2+\cum^2}$ which signifies the
relevance of the Foldy representation.} This suggests that the
operators $\hat{\bfrmx}_{0\{\fa\}}$ and $\hat{\bfrmp}_{0\{\fa\}}$
that clearly satisfy the canonical commutation relations may be
identified with the position and momentum operators acting on the
wave functions $f$. This in turn means that the operators
$\bfX'_0$ and $\bfP'_0$ in the Foldy representation and the
operators $\xa$ and $\pa$ in the $(\cHa,\rh)$-representation also
describe the position and momentum observables. In particular, the
basis vectors $\xi^{(\ep,s)}_\bfx$ and $\Aax^{(\ep,s)}$ determine
the states of the system with a definite position value $\bfx$;
they are {\em localized} in space. They also have definite charge
or chirality (sign of the energy) and spin (say along the
$x^3$-direction).

By construction, $\Aax^{(\ep,s)}$ are delta-function normalized
position eigenvectors, i.e., they satisfy (\ref{orthonormal-A})
and
    \be
    \xa \Aax^{(\ep,s)}=\bfx\: \Aax^{(\ep,s)}.
    \label{x0-A}
    \ee
Similarly, we identify the chirality operator and the spin
operator along the $x^3$-direction with ${\rm
C}=\cUa^{-1}\Sigma_3\cUa$ and $s_{12}=\cUa^{-1}\Sigma_{12}\cUa$,
respectively. This implies
    \be
    {\rm C} \Aax^{(\ep,s)}=\ep \:\Aax^{(\ep,s)}, ~~~~~~~~~
    s_{12} \Aax^{(\ep,s)}=s\: \Aax^{(\ep,s)}.
    \label{C-S-A}
    \ee
In view of Eqs.~(\ref{orthonormal-A}), (\ref{x0-A}), and
(\ref{C-S-A}), the state vectors $\Aax^{(\ep,s)}$ represent
spatially localized Proca fields with definite chirality $\ep$ and
spin $s$. We can associate each Proca field $A\in\cHa$ with a
unique position wave function, namely $f(\ep,s,\bfx)$. As we
explained in Section \ref{phy-obs-sec}, we can use these wave
functions to represent all the physical quantities associated with
the Proca fields. We also emphasize that according to
(\ref{f-helicity}), $f(\ep,s,\bfx)$ are not the helicity
eigenfunctions. Therefore, the state vectors $\Aax^{(\ep,s)}$ do
not have definite helicity. This is in complete accordance with
the Heisenberg uncertainty principle: \emph{since $\Aax^{(\ep,s)}$
are localized states (with definite position) they do not have
definite momentum and hence definite helicity.}

Next, we recall that two Hilbert spaces $\cHa$ for different
choices of the parameters $\{\fa\}$ are unitary-equivalent. We can
use (\ref{ob-4}), (\ref{local}), and the unitary operator
(\ref{tu-A=}) to obtain the explicit form of the localized Proca
fields and the physical observables acting in $\cHa$. The
resulting expressions for the position and spin operators are
highly complicated. Therefore, in the following we shall only
derive the explicit form of these operators in the covariant
representation $(\cH,{\rm h})$.

\subsection{Explicit Form of the Localized States}

As suggested by (\ref{inn-of-f}), the wave functions
$f(\ep,s,\bfx)$ belong to $L^2(\R^3)$. Moreover due to the
implicit dependence of $A^{(\ep,s)}_{\bfx}$ on the $x_0^0$
appearing in the expression for $\cU$, $f(\ep,s,\bfx)$ depend on
$x_0^0$. As in the case of KG fields this dependence becomes
explicit, if we express $f(\ep,s,\bfx)$ in terms of $A$ directly.
To see this, we first substitute (\ref{U-inverse-0}),
(\ref{U-inverse-i}) and (\ref{xi}) in (\ref{local}) to obtain
    \bea
    A^{0(\ep,s^i)}_{\bfx}(x^0)=
    \ep\,\f{D^{-1/4}}{\sqrt{\cum\kappa}}~
     e^{-i\ep(x^0-x^0_0)D^{1/2}}\,\fK^i\,|\bfx\kt,~~~~~
    \bfA^{(\ep,s^i)}_{\bfx}(x^0)=
    \sqrt{\frac{\cum}{\kappa}}~
     \fU^{-1} e^{-i\ep(x^0-x^0_0)D^{1/2}}|\bfx\kt
     \otimes e_{s^i},
    \label{b-1}
    \eea
where $(s^1,s^2,s^3)=(+1,-1,0)$. We then use this equation and
(\ref{inn-inv-2}) to compute the right-hand side of (\ref{f}).
This yields
    \be
    f(\ep,s^i,\bfx)=\sqrt{\frac{\kappa}{\cum}}~
    e^{i\ep(x^0-x^0_0)\hat D^{1/2}}
    \left[\fU\bfA_\ep(x^0,\bfx)\right]^i,
    \label{f2}
    \ee
where $A^\mu_\ep$ is the definite-chirality (definite-energy)
component of $A^\mu$ with chirality $\ep$. Because $A^\mu_\ep$ and
consequently $\fU\bfA_\ep$ satisfy the Foldy equation
(\ref{foldy}), we have $e^{i\ep(x^0-x^0_0) D^{1/2}}\fU
\bfA_\ep(x^0,\bfx)=\fU \bfA_\ep(x_0^0,\bfx)$. Inserting this in
(\ref{f2}), we find the following manifestly $x_0^0$-dependent
expression for $f(\ep,s^i,\bfx)$.
    \be
    f(\ep,s^i,\bfx)=\sqrt{\frac{\kappa}{\cum}}~
    \left[\fU\bfA_\ep(x_0^0,\bfx)\right]^i.
    \label{f3}
    \ee

\begin{figure}
\centerline{\includegraphics[width=.4\columnwidth]{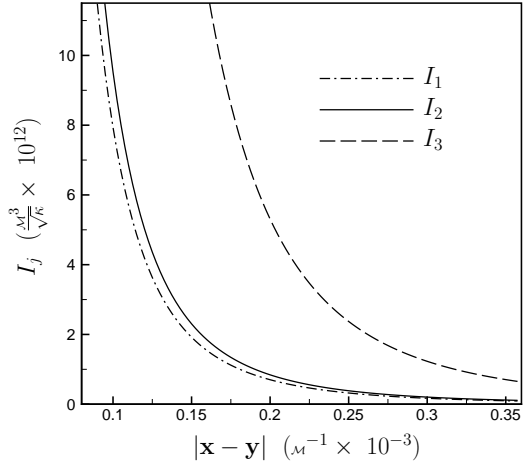}}
\caption{Plots of $I_1, I_2$ and $I_3$ in terms of the radial
distance $|\bfx-\bfy|$. The radial distance is scaled with the
Compton wave length $\cum^{-1}$.} \label{i123}
\end{figure}

Next, we use (\ref{b-1}) to compute the value of the localized
Proca fields $A^{(\ep,s^i)}_{\bfy}$ at a spacetime point
$(x^0,\bfx)$:
    \be
    A^{(\ep,s^i)}_{\bfy}(x^0,\bfx):=\br\bfx|A^{(\ep,s^i)}_{\bfy}(x^0)\kt.
    \ee
Doing the necessary calculations, we obtain
    {\small\bea
    A^{0(\ep,+1)}_{\bfy}(x)\equa
    i\ep\sin\theta\cos\phi\:I_1,~~~~~
    A^{0(\ep,-1)}_{\bfy}(x)=
    i\ep\sin\theta\sin\phi\:I_1,~~~~~
    A^{0(\ep,0)}_{\bfy}(x)=
    i\ep\cos\theta\:I_1,~~~~\label{A0-local}\\
    \bfA^{(\ep,+1)}_{\bfy}(x)\equa\left(\begin{array}{c}
    v_1 \\ v_2 \\ v_3 \end{array}\right),~~~~
    \bfA^{(\ep,-1)}_{\bfy}(x)=\left(\begin{array}{c}
    v_2 \\ v_4 \\ v_5 \end{array}\right),~~~~
    \bfA^{(\ep,0)}_{\bfy}(x)=\left(\begin{array}{c}
    v_3 \\ v_5 \\ v_6 \end{array}\right),\label{vecA-local}
    \eea}
where $x:=(x^0,\bfx)$, $\theta$ and $\phi$ are the polar and
azimuthal angles representing the direction of $\bfx-\bfy$,
    \[\begin{array}{lll}
    v_1:= I_2+\sin^2\theta \cos^2\phi\:I_3, ~~&
    v_2:=\f{1}{2}\sin^2\theta\sin(2\phi)\:I_3, ~~&
    v_3:=\f{1}{2}\sin(2\theta)\cos\phi\:I_3,  \\
    v_4:=I_2+\sin^2\theta\sin^2\phi\:I_3, &
    v_5:=\f{1}{2}\sin(2\theta)\sin\phi\:I_3, &
    v_6:=I_2+\cos^2\theta \:I_3,
    \end{array}\]
and
    \bea
   && I_1:=\f{|\bfx-\bfy|}{2\pi^2\sqrt{{\cum\kappa}}}
    \int_0^\infty\!\!d k\:k^2\:\Omega_1(\ep,k)\:\Omega_2(k),
    \label{I1=z}\\
   && I_2:=\sqrt{\f{\cum}{\kappa}}\:\f{1}{2\pi^2}
    \int_0^\infty\!\!d k\:\Omega_1(\ep,k)\left\{
    \f{k\sin(k|\bfx-\bfy|)}{|\bfx-\bfy|}+
    \Omega_2(k)\left\{\cum^{-1}[k^2+\cum^2]^{1/2}-1
    \right\}\right\},\\
   && I_3:=\sqrt{\f{\cum}{\kappa}}\:\f{1}{2\pi^2}
    \int_0^\infty\!\!d k\:\Omega_1(\ep,k)\left\{
    \f{k\sin(k|\bfx-\bfy|)}{|\bfx-\bfy|}-3\,\Omega_2(k)\right\}
    \left\{\cum^{-1}[k^2+\cum^2]^{1/2}-1\right\},
    \label{I3=z}\\
   && \Omega_1(\ep,k):=\f{\exp{\left[-i\ep(x^0-x^0_0)
    \sqrt{k^2+\cum^2}\:\right]}}{
    [k^2+\cum^2]^{1/4}},~~~~~~
    \Omega_2(k):=\f{\sin(k|\bfx-\bfy|)}{k|\bfx-\bfy|^3}-
    \f{\cos(k|\bfx-\bfy|)}{|\bfx-\bfy|^2}.~~~~~~
    \eea
For $x^0=x_0^0$, the integrals on the right-hand side (\ref{I1=z})
-- (\ref{I3=z}) can be expressed in terms of the Bessel K-function
($K_n$), and the Hypergeometric function ($_pF_q$). The result,
for both $\epsilon=-1$ and $1$, is
    {\small\bea
    I_1\equa\sqrt{\f{\cum}{\kappa}}~\left[2^\frac{3}{4}\pi^\frac{3}{2}
    \Gamma(\mbox{\footnotesize$\frac{1}{4}$})\right]^{-1}
    \left(\frac{\cum}{z}\right)^\frac{5}{4}
    \left\{\f{5}{2\cum z}\,K_\frac{5}{4}(\cum z)+
    K_\frac{1}{4}(\cum z)\right\},
    \label{exn2}\\
    I_2\equa\sqrt{\f{\cum}{\kappa}}~\left[2^\frac{3}{4}\pi^\frac{3}{2}
    \Gamma(\mbox{\footnotesize$\frac{1}{4}$})\right]^{-1}
    \left(\frac{\cum}{z}\right)^\frac{5}{4}
    \left\{K_\frac{5}{4}(\cum z)+\f{1}{\cum z} K_\frac{1}{4}(\cum
    z)+\f{\Gamma(\mbox{\footnotesize$\frac{1}{4}$})^2}{4\pi(\cum z)^\f{3}{2}}
    K_\frac{3}{4}(\cum z)+\right.\nn\\
    & &\vspace{2cm}\hspace{0cm}\left.\f{1}{(\cum
    z)^\f{3}{4}}\left(\f{2\pi}{\Gamma(\mbox{\footnotesize$\frac{1}{4}$})}
    ~_1\!F_2\left[\f{1}{2};\f{5}{4},\f{3}{2};\f{\cum^2 z^2}{4}\right]
    +\f{\Gamma(\mbox{\footnotesize$\frac{1}{4}$})^3}{3\pi.2^\f{7}{4}}
    ~_1\!F_2\left[\f{1}{2};\f{3}{2},\f{7}{4};\f{\cum^2z^2}{4}\right]\right)
    +\right.\nn\\
    & &\vspace{2cm}\hspace{0cm}\left.
    \f{\Gamma(\mbox{\footnotesize$\frac{1}{4}$})}{(\cum z)^\f{5}{4}}
    \left(\f{1}{2^\f{3}{4}\cum z}
    ~_1\!F_2\left[\f{1}{4};\f{1}{4},\f{3}{4};\f{\cum^2 z^2}{4}\right]
    -2^\f{1}{4}~_1\!F_2\left[\f{1}{4};\f{3}{4},\f{5}{4};\f{\cum^2z^2}{4}\right]\right)
    \right\},\\
    I_3\equa\sqrt{\f{\cum}{\kappa}}~\left[2^\frac{3}{4}\pi^\frac{3}{2}
    \Gamma(\mbox{\footnotesize$\frac{1}{4}$})\right]^{-1}
    \left(\frac{\cum}{z}\right)^\frac{5}{4}
    \left\{\f{3K_\frac{1}{4}(\cum z)}{\cum z}+
    \f{K_\frac{5}{4}(\cum z)}{(\cum z)^\f{5}{4}}
    +\f{\Gamma(\mbox{\footnotesize$\frac{1}{4}$})^2}{4\pi(\cum z)^\f{1}{2}}
    \left[K_\frac{7}{4}(\cum z)+\f{3K_\frac{3}{4}(\cum z)}{\cum z}\right]
    +\right.\nn\\
    & &\vspace{2cm}\hspace{0cm}\left.\f{1}{(2\cum z)^\f{3}{4}}
    \left(\f{12\pi}{\Gamma(\mbox{\footnotesize$\frac{1}{4}$})}
    ~_1\!F_2\left[\f{1}{2};\f{5}{4},\f{3}{2};\f{\cum^2 z^2}{4}\right]
    +\f{\Gamma(\mbox{\footnotesize$\frac{1}{4}$})^3}{2\pi}
    ~_1\!F_2\left[\f{1}{2};\f{3}{2},\f{7}{4};\f{\cum^2 z^2}{4}\right]\right)
    +\right.\nn\\
    & &\vspace{2cm}\hspace{0cm}\left.
    \f{3\Gamma(\mbox{\footnotesize$\frac{1}{4}$})}{(\cum z)^\f{5}{4}}
    \left(\f{1}{2^\f{3}{4}\cum z}
    ~_1\!F_2\left[-\f{1}{4};\f{1}{4},\f{3}{4};\f{\cum^2 z^2}{4}\right]
    +2^\f{1}{4}~_1\!F_2\left[\f{1}{4};\f{3}{4},\f{5}{4};\f{\cum^2 z^2}{4}\right]\right)
    \right\},
    \eea}%
where $z:=|\bfx-\bfy|$ and $\Gamma$ is the Gamma function.
Fig.~\ref{i123} gives the graphs of $I_1,I_2,$ and $I_3$. They
involve a $\delta$-function-like singularity at $|\bfx-\bfy|=0$.
This is a manifestation of the fact that $A^{(\ep,s)}_{\bfy}$ are
localized at $\bfy$.

As seen from Eqs.~(\ref{A0-local}) and (\ref{vecA-local}), the
localized Proca fields, unlike the localized KG fields
\cite{MZ-ann06-1}, depend on the angles $\theta$ and $\phi$
(direction of $\bfx-\bfy$). But as for the KG fields their
position wave functions involve the delta functions; the position
wave function $f_{(\ep,s,\bfx)}(\ep',s',\bfx')$ for
$A^{(\ep,s)}_\bfx(x_0^0)$ has the form
$\delta_{\ep,\ep'}\delta_{s,s'}\delta^3(\bfx-\bfx')$.

Finally, we wish to emphasize that to the best of our knowledge
the explicit form of the localized Proca fields
$A^{(\ep,s)}_{\bfx}$ have not been previously given. It is
remarkable that we have obtained these localized states without
pursuing the axiomatic approach of Ref.~\cite{ne-wi}. The latter
gives rise to the Bargmann-Wigner localized fields
\cite{ne-wi,azc}.

\subsection{Position, Spin, and Helicity Operators}

Using the unitary map $\cU$ and its inverse, we can obtain the
explicit form of the position operator $\bfrmx_0$ that is defined
to act on the Proca fields $A\in\cH$. Note that $\chi:=\bfrmx_0A$
is a three-component field whose components satisfy both
(\ref{A-d}) and (\ref{Lorentz-condition-HS}). Therefore it is
uniquely determined in terms of the initial data
$(\chi(x^0_0),\dot{\chi}(x^0_0))$ for some $x^0_0\in\R$. One can
compute the latter using (\ref{ob-4}), (\ref{u-A=}),
(\ref{U-inverse-0}), (\ref{U-inverse-i}), and the identities:
$D=\bffK^2+\cum^2$,
    \bea
    \fh \bfS \fh\equa\f{\bffK}{|\bffK|}\,\fh,
    \label{hel-props}\\
    \left[\bfrmx,\fh\right]\equa\f{i}{|\bffK|}\,\bfS -
    \f{i\bffK}{|\bffK|^2}\,\fh,~~~~~~~~
    \left[\bfrmx,\fh^2\right]=\f{i}{|\bffK|}\,
    (\fh\bfS+\bfS \fh) - \f{2i\bffK}{|\bffK|^2}\,\fh^2,
    \label{x-hel-com}
    \eea
and $[F(\bffK),\bfrmx]=-i\bfdel_\bffK F(\bffK)$, where $F$ is a
differentiable function. This yields
    {\small\bea
    \chi^0(x^0_0)\equa\bffY A^0(x^0_0)+
    \f{1}{\cum D^{1/2}}\,\dot\bfA(x^0_0),~~~~~~~~~
    \dot\chi^0(x^0_0)=\left(\bffY-
    \f{i\bffK}{D}\right)\dot{A}^0(x^0_0)-
    \f{D^{1/2}}{\cum}\,\bfA(x^0_0),
    \label{x-A-0}\\
    \bfchi(x^0_0)\equa\bffX\,\bfA(x^0_0),~~~~~~~~~~~
    \dot{\bfchi}(x^0_0)=\left(\bffX-
    \f{i\bffK}{D}\right)\dot\bfA(x^0_0),
    \label{x-A}
    \eea}
where
    \bea
    \bffY\equdef\bfrmx+\f{i\bffK}{2D}+\f{i\bffK}{\cum(D^{1/2}+
    \cum)},\\
    \bffX\equdef\bfrmx-\f{i}{2}\,\bffK D^{-1}+
    i\bffK[D^{-1}-\cum^{-1}D^{-1/2}]\fh^2+
    \f{i}{|\bffK|}[1-\cum D^{-1/2}]\fh\bfS+
    \f{i}{|\bffK|}[\cum^{-1}D^{1/2}-1]\bfS\fh, \nn\\
    \equa \bfrmx-\f{i\bffK}{2D}-\f{i\bffK(\bffK\cdot\bfS)^2}
    {\cum D(D^{1/2}+\cum)}+
    \f{i\left[\bfS(\bffK\cdot\bfS)+(\bffK\cdot\bfS)
    \bfS\right]}{2\cum D^{1/2}}-
    \f{(D^{1/2}-\cum)\,(\bffK\times\bfS)}{2\cum D^{1/2}
    (D^{1/2}+\cum)}.
    \eea

Next, we employ (\ref{g-A-sol}) to express $\chi(x^0)$ in terms of
(\ref{x-A-0}) and (\ref{x-A}). Simplify the resulting expression,
we find, for all $x^0\in\R$,
    \bea
    \chi^0(x^0)\equa\left\{\bffY-i(x^0-x^0_0)\bffK\, D^{-1}\p_0
    \right\}A^0(x^0)+
    \f{1}{\cum D^{1/2}}\dot\bfA(x^0),\label{chi-scalar}\\
    \bfchi(x^0)\equa\left\{\bffX-i(x^0-x^0_0)\bffK\, D^{-1}\p_0\right\}\bfA(x^0).
    \label{chi-vector}
    \eea
In addition to being a Hermitian operator acting in the physical
Hilbert space $\cH$, the position operator $\bfrmx_0$ has the
following notable properties.
    \begin{enumerate}
    \item It respects the charge superselection rule
    \cite{superselection}, for it commutes with the
    chirality operator ${\rm C}=s_3$. This is easily seen
    by noting that $\bfrmx_0$ and $s_3$ are  respectively obtained
    via a similarity transformation (\ref{ob-4}) from
    $\bfX'_0$ and $\cS'_3=\Sigma_3$, and that according to
    (\ref{ob-0}), $[\bfX'_0,\Sigma_3]=0$.
    \item It has commuting components, and it
    commutes with the spin operator. This is because it is obtained
    via a similarity transformation (\ref{ob-4}) from $\bfX'_0$
    which has commuting components and commutes with $\bfS'$.
    \item Its $x^0$-derivative gives the relativistic velocity
    operator; i.e., we have:
        \[\mathbf{v}:=\f{d\bfrmx_0}{d x^0}=
        i\left[\rh,\bfrmx_0\right]=
        i\cU^{-1}\left[H',\bfX'\right]\cU=
        \f{\bfrmp_0}{\sqrt{\bfrmp_0^2+m^2c^2}}\,{\rm C}.\]
    As we noticed in \cite{MZ-ann06-2}, this result means that the physical momentum is
    $\bfrmp_0\,{\rm C}$.
    \item It has the correct nonrelativistic limit: As
    $c\to\infty$, $\bfrmx_0\to \bfrmx$.
    \end{enumerate}

Similarly, we can obtain the explicit form of the spin operator
$\bfs_0$ acting on $A\in\cH$. Again, noting that
$\mathscr{S}:=\bfs_0 A$ is a three-component field whose
components satisfy (\ref{A-d}) and (\ref{Lorentz-condition-HS}),
we can determine $\mathscr{S}$ in terms of the initial data
$(\mathscr{S}(x^0_0),\dot{\mathscr{S}}(x^0_0))$. Using
(\ref{u-A=}), (\ref{U-inverse-0}), (\ref{U-inverse-i}),
(\ref{ob-4}), (\ref{hel-props}), and
    \be
    \left[\bfS,\fh\right]=\f{i}{|\bffK|}\,(\bffK\times\bfS),~~~~~~~~~
    \left[\bfS,\fh^2\right]=\f{i}{|\bffK|}
    \left[(\bffK\times\bfS)\fh+
    \fh(\bffK\times\bfS)\right],
    \label{S-hel-com}
    \ee
these initial values are given by
    \be
    \begin{array}{ll}
    \mathscr{S}^0(x^0_0)=\cum^{-1} D^{-1/2}\bffK
    \times\dot\bfA(x^0_0),~~~~ &
    \dot{\mathscr{S}}^0(x^0_0)=-\cum^{-1} D^{1/2}\bffK
    \times\bfA(x^0_0),\\
    \vec{\mathscr{S}}(x^0_0)=\tilde{\bfs}_0\bfA(x^0_0), &
    \dot{\vec{\mathscr{S}}}(x^0_0)=\tilde{\bfs}_0\dot\bfA(x^0_0),
    \end{array}
    \label{s-A}
    \ee
where
    \bea
    \tilde{\bfs}_0\equa\bfS+\f{i}{|\bffK|}
    [1-\cum D^{-1/2}]\fh(\bffK\times\bfS)+
    \f{i}{|\bffK|}[\cum^{-1} D^{1/2}-1](\bffK\times\bfS)\fh\nn\\
    \equa\f{D+\cum^2}{2\cum D^{1/2}}\,\bfS-
    \f{(D^{1/2}-\cum)\,\bffK (\bffK\cdot\bfS)}{2\cum D^{1/2}(D^{1/2}+\cum)}+
    \f{i\left\{(\bffK\cdot\bfS)(\bffK\times\bfS)+
    (\bffK\times\bfS)(\bffK\cdot\bfS)\right\}}{2\cum D^{1/2}}.
    \eea
Next, we use (\ref{g-A-sol}) to express $\mathscr{S}(x^0)$ in
terms of the initial data (\ref{s-A}). This leads to
    \be
    \mathscr{S}^0(x^0)=\cum^{-1} D^{-1/2}\bffK\times\dot\bfA(x^0),~~~~~~~~~
    \vec{\mathscr{S}}(x^0)=\tilde{\bfs}_0\bfA(x^0).
    \label{s-A-t}
    \ee
As seen from (\ref{chi-scalar}), (\ref{chi-vector}), and
(\ref{s-A-t}) the action of the position and spin operators on a
Proca field $A$ mixes the components of the field. It is clearly
more complicated than the action of the corresponding operators
for the KG fields \cite{MZ-ann06-1}.

Next, we evaluate the action of the momentum, angular momentum,
and helicity operators on $A$. Because $\bfP'_0$ and $\rho$
commute, in view of (\ref{ob-4}) and (\ref{U=}), we have
$\bfrmp_0=U_{x^0_0}^{-1}\bfP'_0 U_{x^0_0}$. This in turn implies
$[\bfrmp_0\,A](x^0)=\bfrmp\,A(x^0)$ for all $x^0\in\R$.
Furthermore, using (\ref{chi-scalar}), (\ref{chi-vector}), and
(\ref{s-A-t}), we can show that the angular momentum operator (in
units of $\hbar$) $\bfL:=\hbar^{-1}\bfrmx_0\times\bfrmp_0$ acts on
$A$ according to
    \be
    \begin{array}{l}
    \mathscr{L}^0(x^0)=\hbar^{-1}(\bfrmx\times\bfrmp) A^0(x^0)-\mathscr{S}^0(x^0), \\
    \vec{\mathscr{L}}(x^0)=\left\{
    \hbar^{-1}\bfrmx\times\bfrmp+\bfS\right\}A(x^0)
    -\vec{\mathscr{S}}(x^0), \\
    \end{array}~~~~~~~~\forall x^0\in\R,
    \ee
where $\mathscr{L}(x^0):=[\bfL A](x^0)$, and
$\mathscr{S}(x^0):=[\bfs_0 A](x^0)$ is given in (\ref{s-A-t}).
Therefore, unlike the position $\bfrmx_0$ and spin $\bfs_0$
operators, the (linear) momentum $\bfrmp_0$ and the total angular
momentum $\mathbf{M}:=\bfL+\bfs_0$ operators have the same
expressions as in nonrelativistic QM.\footnote{Note that the
corresponding particle in nonrelativistic QM is described by
$\bfA$ \cite{Yndurain}, i.e., we should set $A^0=0$.} Similarly,
we can compute the action of the helicity operator on $A$. In view
of (\ref{s-A-t}), we find:
$[(\bfrmp_0\cdot\bfs_0)\,A](x^0)=\big(0,(\bfp\cdot\bfS)\bfA(x^0)\big)$,
which is in complete accordance with our previous result
(\ref{helicity-A-exp}).

Again, to the best of our knowledge, the actions of the position,
spin, and other observables on a (covariant) Proca field $A$ have
not been previously given. Earlier works on the subject
\cite{other-position,azc,ne-wi,localizations} calculate the
position and spin operators that act either on the six-component
fields, i.e., in the Hilbert space $\cK$, or on the
Bargmann-Wigner's second-rank 4-spinors. One can use the unitary
transformation $\rho:\cK\to\cH'$ to compute the position
($\bfX_0:=\rho^{-1}\bfX'_0\rho$) and spin
($\bfS_0:=\rho^{-1}\bfS'\rho$) operators acting in $\cK$. The
result is
    {\small\bea
    \bfX_0\equa\bfrmx-\f{D^{1/2}-\cum}
    {2\cum D^{1/2}(D^{1/2}+\cum)}\,(\bffK\times\bfS)+
    \sigma_1\left\{-\,\f{i\bffK}{2D}-\f{i\bffK(\bffK\cdot\bfS)^2}
    {\cum D(D^{1/2}+\cum)}+
    \f{i\,\left[\bfS(\bffK\cdot\bfS)+(\bffK\cdot\bfS)
    \bfS\right]}{2\cum D^{1/2}}\,
    \right\},~~~~~\\
    \bfS_0\equa\f{D+\cum^2}{2\cum D^{1/2}}\,\bfS-
    \f{D^{1/2}-\cum}{2\cum D^{1/2}(D^{1/2}+\cum)}\,
    \bffK (\bffK\cdot\bfS)+
    \f{i}{2\cum D^{1/2}}\,\sigma_1
    \left\{(\bffK\cdot\bfS)(\bffK\times\bfS)+
    (\bffK\times\bfS)(\bffK\cdot\bfS)\right\}.~~~~~
    \eea}
These are exactly the position and spin operators that are
obtained by Case in \cite{Case}.

Following the treatment of the KG fields in
\cite{M-IJMPA,MZ-ann06-2} we can identify the coherent states of
Proca fields with the eigenstates of the annihilation operator
$\bfa:=\sqrt{\frac{k}{2\hbar}}\,
\left(\bfrmx_0+ik^{-1}\bfrmp_0\right)$, where $k\in\R$. Because
both $\bfrmx_0$ and $\bfrmp_0$ commute with the chirality ($s_3$)
and spin ($s_{12}$) operators, so does $\bfa$. Hence, we can
introduce a set of coherent states with definite chirality and
spin. The corresponding state vectors $|\bfz,\ep,s)$ are defined
as the common eigenvectors of $\bfa$, $s_3$ and $s_{12}$, i.e.,
$\bfa|\bfz,\ep,s)=\bfz|\bfz,\ep,s)$ and
$s_3|\bfz,\ep,s)=\ep|\bfz,\ep,s)$,
$s_{12}|\bfz,\ep,s)=s|\bfz,\ep,s)$, where $\bfz\in\C^3$,
$\ep\in\{-,+\}$ and $s\in\{-1,0,+1\}$. We can studied these
coherent states and found that they have essentially the same
properties as the coherent states of the KG fields that we
explored in \cite{MZ-ann06-2}.

\section{Probability Density for Spatial Localization of a Field}

We may employ the procedure outlined in \cite{MZ-ann06-1} to find
the probability density for the spatial localization of a Proca
field. As in nonrelativistic QM, we identify the probability of
the localization of a Proca field $A$ in a region
$V\subseteq\R^3$, at time $t_0=x_0^0/c$, with
    \be
    P_V=\int_V d^3\bfx~\|\Pi_{\bfx}A\|^2,
    \label{project-proca}
    \ee
where $\Pi_{\bfx}$ is the projection operator onto the eigenspace
of $\bfrmx_0$ with eigenvalue $\bfx$, i.e.,
    \be
    \Pi_{\bfx}=\sum_{\ep=\pm}\sum_{s=0,\pm1}
    |A^{(\ep,s)}_\bfx)(A^{(\ep,s)}_\bfx|,
    \label{prob-subst-z}
    \ee
$\|\cdot\|^2:=\cbr\cdot,\cdot\ckt$ is the square of the norm of
${\cal H}$, and we assume $\|A\|=1$. Substituting
(\ref{prob-subst-z}) in (\ref{project-proca}) and making use of
(\ref{orthonormal-A}) and (\ref{f}), we have
$P_V=\sum_{\ep=\pm}\sum_{s=0,\pm1}\int_V
d^3\bfx~|f(\ep,s,\bfx)|^2$. Therefore, in light of (\ref{f3}),
(\ref{A-c}), and (\ref{A-pm-field}), the probability density is
given by
    \be
    \varrho(x_0^0,\bfx):=\sum_{\ep=\pm}\sum_{s=0,\pm1}
    |f(\ep,s,\bfx)|^2=\frac{\kappa}{2\cum}
    \left\{|\fU\bfA(x^0_0,\bfx)|^2+
    |\fU\,D^{-1/2}\dot\bfA(x^0_0,\bfx)|^2\right\}.
    \label{probd=}
    \ee
For a position measurement to be made at time $t=x^0/c$, we have
the probability density
    \be
    \varrho(x^0,\bfx)=\frac{\kappa}{2\cum}\left\{
    |\fU\bfA(x^0,\bfx)|^2+
    |\fU\,D^{-1/2}\dot\bfA(x^0,\bfx)|^2\right\}
    =\frac{\kappa}{2\cum}\left\{
    |\fU\bfA(x)|^2+|\fU\bfA_c(x)|^2\right\}.
    \label{probd}
    \ee

We can use the method discussed in Section~\ref{cons-current-sec}
to introduce a current density ${\cal J}^\mu$ such that ${\cal
J}^0=\varrho$ (See also \cite{rosenstein-horwitz}.) This
probability current density turns out to have the form
    \be
    {\cal J}^\mu(x)=\f{\kappa}{2\cum}\Re\left\{
    \fU\bfA(x)^*\cdot\left(D^{-1}\p^\mu\fU\dot\bfA(x)\right)+
    \fU\bfA_c(x)^*\cdot\left(D^{-1}\p^\mu\fU\dot\bfA_c(x)\right)
    \right\}+\Upsilon^\mu(x),~~~~
    \label{cal-J0}
    \ee
where $\Re$ means ``the real part of", $\Upsilon^0(x)=0$ and, for
all $i\in\{1,2,3\}$,
    \bea
    \Upsilon^i(x)\equa\f{\kappa}{2\cum}\Re\left\{
    \left(\fU\bfA(x)\right)^{i*}
    \left[\bfdel\cdot\f{D^{-1/2}}{D^{1/2}+\cum}\fU\dot\bfA(x)\right]-
    \left[\fU\bfA(x)^{i*}\cdot\bfdel\right]
    \left(\f{D^{-1/2}}{D^{1/2}+\cum}\fU\dot\bfA(x)\right)^i+\right.\nn\\
    &&\vspace{2cm}\hspace{0cm}\left.
    \left(\fU\bfA_c(x)\right)^{i*}
    \left[\bfdel\cdot\f{D^{-1/2}}{D^{1/2}+\cum}\fU\dot\bfA_c(x)\right]-
    \left[\fU\bfA_c(x)^{i*}\cdot\bfdel\right]
    \left(\f{D^{-1/2}}{D^{1/2}+\cum}\fU\dot\bfA_c(x)\right)^i\right\}\!.~~~
    \eea
One can easily show that ${\cal J}^\mu(x)$ is neither a
four-vector nor a conserved current density.

Although the above discussion is based on a particular choice for
the parameters $\{\fa\}$, namely $\{\fa\}=\{1\}$, it is generally
valid. This is because the position wave functions $f$ and the
corresponding probability densities do not depend on the
parameters ${\{\fa\}}$. Therefore, if we are to compute the
probability density $\vrhoa$ of the spatial localization of a
Proca field $A\in\cHa$ with the position operator being identified
with $\xa$ for $\{\fa\}\neq\{1\}$, we have, for a measurement made
at $t_0=x_0^0/c$,
    \be
    \vrhoa(x_0^0,\bfx)=\f{\kappa}{2\cum}\left\{|\fU\,\bfA'_{\{\fa\}}(x^0_0,\bfx)|^2+
    |\fU\,D^{-1/2}\dot\bfA'_{\{\fa\}}(x^0_0,\bfx)|^2\right\},
    \label{rho-a}
    \ee
where $\bfA'_{\{\fa\}}:=\Ua^{-1}\bfA$ and $\Ua:\cH\to\cHa$ is
given by (\ref{cur-u-a}). We can compute $\bfA'_{\{\fa\}}$ using
(\ref{u-A=}), (\ref{tU-inverse-i}), and (\ref{cur-u-a}) and use
the result to obtain
    \bea
    \vrhoa(x)\equa\frac{\kappa}{4\cum}\left\{
    |\fU_{+,+}\bfA(x)|^2+|\fU_{-,+}\bfA(x)|^2+
    |\fU_{+,+}\bfA_c(x)|^2+|\fU_{-,+}\bfA_c(x)|^2\right.\nn\\
    &&\vspace{2cm}\left.+2\Re\left[
    (\fU_{+,+}\bfA(x))^*\cdot\fU_{+,+}\bfA_c(x)-
    (\fU_{-,+}\bfA(x))^*\cdot\fU_{-,+}\bfA_c(x)\right]\right\}.
    \label{rho-a-3}
    \eea
Again we can use the method of Section~\ref{cons-current-sec} to
compute a probability current density ${\cal J}\suba^\mu$ such
that ${\cal J}\suba^0=\vrhoa$. This leads to a complicated
expression that we do not include here. Similarly to ${\cal
J}^\mu$, we expect ${\cal J}\suba^\mu$ to be neither covariant nor
conserved.

The non-conservation (respectively non-covariance) of the
probability current density ${\cal J}\suba^\mu$ raises the issue
of the non-conservation (respectively frame-dependence) of the
total probability:
    \be
    \fPa:=\int_{\R^3}d^3\bfx~\vrhoa(x^0,\bfx).
    \label{Prob-00}
    \ee
This would certainly be unacceptable. The situation is analogous
to that of the KG fields. $\fPa$ is indeed a frame-independent
conserved quantity, thanks to the covariance and conservation of
the current density $J\suba^\mu$ and the identity
    \be
    \int_{\R^3}d^3\bfx~\vrhoa(x^0,\bfx)=
    \int_{\R^3}d^3\bfx~J\suba^0(x^0,\bfx),
    \label{Prob-01}
    \ee
which follows from (\ref{j-zero-gen}), (\ref{rho-a-3}) and the
fact that $\fU_{\ep,\ep'}$ are self-adjoint operators acting in
$\tcH$.

Combining (\ref{Prob-00}) and (\ref{Prob-01}), we have
    \be
    \fPa=\int_{\R^3}d^3\bfx~J\suba^0(x^0,\bfx).
    \label{rho=j}
    \ee
This relation implies that although the probability density
$\vrhoa$ is not the zero-component of a conserved four-vector
current density, its integral over the whole space that yields the
total probability (\ref{Prob-00}) is nevertheless conserved.
Furthermore, this global conservation law stems from a local
conservation law, i.e., a continuity equation for a four-vector
current density namely $J\suba^\mu$.

\section{Gauge Symmetry Associated with the Conservation of the
Total Probability}

In this section we explore a global gauge symmetry that supports
the conservation of the total probability or its local realization
as the conservation of the current density $J\suba^\mu$. To
determine the nature of this symmetry, we recall that the
conserved charge associated with any conserved current is the
generator of the corresponding infinitesimal gauge transformations
\cite{itzykson-zuber}. We use the Hamiltonian formulation to
obtain these transformations. The procedure we follow mimics the
one presented in \cite{MZ-ann06-1} for the KG fields.

The Lagrangian $L$ for a Proca field $A$ and the corresponding
canonical momenta $\Pi(\bfx)$, $\bar\Pi(\bfx)$ associated with
$A(\bfx):=A(x^0,\bfx)$ and $A^*(x^0,\bfx)$ are respectively given
by \cite{weinberg}: \footnote{Throughout this section we suppress
the $x^0$-dependence of the fields for simplicity.}
    \bea
    L\equdef-\int_{\R^3}d^3\bfx\left\{
    \f{1}{2} F_{\mu\nu}(\bfx)^* F^{\mu\nu}(\bfx)+
    \cum^2 A_\mu(\bfx)^* A^\mu(\bfx)\right\},
    \label{Lag}\\
    \Pi^0(\bfx)\equdef\frac{\delta L}{\delta\dot A_0(\bfx)}=0,
    ~~~~~~\Pi^i(\bfx):=\frac{\delta L}{\delta{\dot A}_i(\bfx)}=
    -F^{0i}(\bfx)^*=-E^i(\bfx)^*,
    \label{mom}\\
    \bar\Pi^0(\bfx)\equdef\frac{\delta L}{\delta\dot{A}_0(\bfx)^*}=0,
    ~~~~~~\bar\Pi^i(\bfx):=\frac{\delta L}{\delta\dot{A}_i(\bfx)^*}=
    -F^{0i}(\bfx)=-E^i(\bfx).
    \label{mom-star}
    \eea
The fact that $\Pi^0$ and $\bar\Pi^0$ vanish show that the Proca
system is a constrained system. There are two primary constraints:
\footnote{Following Dirac's notation \cite{Dirac}, we write the
constraints as \emph{weak equations} with the \emph{weak equality
symbol} `$\approx$'.}
    \be
    \Phi_1:=\Pi^0(x)\approx 0,~~~~~~~~~~
    \Phi_2:=\bar\Pi^0(x)\approx 0.
    \label{primary}
    \ee
Solving for the velocities $\dot A_i$ and $\dot{A}_i^*$ in
(\ref{mom}) and (\ref{mom-star}), and using the Hamiltonian
    \bea
    H_0\equdef\int_{\R^3}\!\!d^3\bfx\left\{
    \Pi_i(\bfx)\bar\Pi^i(\bfx)-\Pi^i(\bfx)\p_iA_0(\bfx)
    -\bar\Pi^i(\bfx)\p_i A_0(\bfx)^*+
    \cum^2A_0(\bfx) A^0(\bfx)^*+\right.\nn\\
    &&\hspace{3cm}\left.\cum^2A_i(\bfx) A^i(\bfx)^*
    +\p_i A_j(\bfx)\p^i A^j(\bfx)^*-
    \p_i A_j(\bfx)\p^j A^i(\bfx)^*\right\},
    \label{fc-H}
    \eea
we obtain the so-called \emph{total} Hamiltonian \cite{Dirac}:
$H_T:=H_0+u^j\Phi_j$, where $u^j, j=1,2$ are two unknown
coefficients. Using the Poisson bracket
    \be
    \left\{\cF,\cG\right\}_P:=
    \int_{\R^3}\!\!d^3\bfx\left[
    \frac{\delta\cF}{\delta A_\mu(\bfx)}
    \frac{\delta\cG}{\delta\Pi^\mu(\bfx)}
    -\frac{\delta\cG}{\delta A_\mu(\bfx)}
    \frac{\delta\cF}{\delta\Pi^\mu(\bfx)}
    +\frac{\delta\cF}{\delta{A}_\mu(\bfx)^*}
    \frac{\delta\cG}{\delta{\bar\Pi}^\mu(\bfx)}
    -\frac{\delta\cG}{\delta{A}_\mu(\bfx)^*}
    \frac{\delta\cF}{\delta{\bar\Pi}^\mu(\bfx)}\right],
    \label{poisson}
    \ee
of the observables $\cF$ and $\cG$, we can easily show that the
dynamical consistency of the primary constraints (\ref{primary}),
i.e., $\dot\Phi_j=\left\{\Phi_j,H_T\right\}_P\approx 0$, results
in the following secondary constraints,
    \be
    \Phi_3:=\cum^2 A_0(\bfx)+\p_i\bar\Pi^i(\bfx)\approx 0,
    ~~~~~~~~~~~
    \Phi_4:=\cum^2 A_0(\bfx)^*+\p_i\Pi^i(\bfx)\approx 0,
    \label{secondary}
    \ee
and that there is no other secondary constraint.

It is not difficult to show that the matrix $\scrC_{j
j'}(\bfy,\bfz):=\left\{\Phi_j(\bfy),\Phi_{j'}(\bfz)\right\}_P$ is
nonsingular. Hence we have a theory with four \emph{second-class}
constraints and can apply Dirac's canonical quantization that uses
the Dirac bracket \cite{Dirac}:
    \be
    \left\{\cF,\cG\right\}_D:=
    \left\{\cF,\cG\right\}_P-
    \int_{\R^3}\!\!d^3\bfy\int_{\R^3}\!\!d^3\bfz
    \left\{\cF,\Phi_j(\bfy)\right\}_P
    \scrC^{j j'}(\bfy,\bfz)\left\{\Phi_{j'}(\bfz),\cG\right\}_P,
    \label{dirac-br}
    \ee
where $\scrC^{j j'}(\bfy,\bfz)$ is the inverse of $\scrC_{j
j'}(\bfy,\bfz)$. Computing the latter and its inverse, we find
$\scrC^{j j'}(\bfy,\bfz):=\tilde{\scrC}^{j
j'}\delta^3(\bfy-\bfz)$, and
    \be
    \left\{\cF,\cG\right\}_D=
    \left\{\cF,\cG\right\}_P-
    \int_{\R^3}\!\!d^3\bfy
    \left\{\cF,\Phi_j(\bfy)\right\}_P
    \tilde{\scrC}^{j j'}\left\{\Phi_{j'}(\bfy),\cG\right\}_P,
    \label{dirac-br-2}
    \ee
where
$\tilde{\scrC}^{13}=\tilde{\scrC}^{24}=-\tilde{\scrC}^{31}=-\tilde{\scrC}^{42}=\cum^{-2}$,
and $\tilde{\scrC}^{j j'}=0$ for other $j$'s and $j'$'s. Further
details of the constraint quantization of Proca system can be
found in \cite{Gitman-Henneaux,KimPark,SJZ}.

In terms of the canonical phase space variables $(A,\Pi)$ and
$(A^*,\bar\Pi)$, the total probability~(\ref{rho=j}) takes the
form
    \bea
    \fPa\equa\f{\kappa}{2\cum}\int_{\R^3}d^3\bfx\:\left\{
    \Pi^i(\bfx)\left[\Theta_{+,0}D^{-1/2}\dot\bfA(\bfx)\right]_i-
    A_i(\bfx)^*\left[\Theta_{+,0}D^{-1/2}\dot{\bar\bfPi}(\bfx)\right]^i
    \right.\nn\\&&\vspace{3cm}\hspace{3.75cm}\left.+i\left(
    \Pi^i(\bfx)\left[\Theta_{+,0}\bfA(\bfx)\right]_i-
    A_i(\bfx)^*\left[\Theta_{+,0}\bar\bfPi(\bfx)\right]^i \right)\right\},
    \label{prob=}
    \eea
where we have made use of (\ref{j-zero-gen}), (\ref{mom}), and
(\ref{mom-star}). Now, we can obtain the infinitesimal symmetry
transformation,
    \be
    A\to A+\delta A,
    \label{trans}
    \ee
generated by $\fPa$ using
    \be
    \delta A(\bfx)=\left\{A(\bfx),\fPa\right\}_D
    \delta\phi,
    \label{var}
    \ee
where $\delta\phi$ is an infinitesimal real parameter. In view of
(\ref{mom}), (\ref{mom-star}), (\ref{poisson}), (\ref{dirac-br-2})
-- (\ref{var}), (\ref{maxeq}), (\ref{div-grad}), and (\ref{A-c}),
we have
    \bea
    \delta A^0(\bfx)\equa-i\delta\theta\left\{
    L^0_+{\rm C}+L^0_-\right\}A^0(\bfx),
    \label{gauge-scal}\\
    \delta\bfA(\bfx)\equa-i\delta\theta\left\{
    \Theta_{+,0} {\rm C}+\Theta_{-,0}\right\}\bfA(\bfx),~~~~~~
    \label{gauge-vec}
    \eea
where $\delta\theta:=\kappa\,\delta\phi/(2\cum)$ and
$\Theta_{\pm,0}$ are given by (\ref{theta-pm0}). We may employ
(\ref{helicity-A-exp}) to express (\ref{gauge-scal}) and
(\ref{gauge-vec}) as
    \be
    \delta A(\bfx)=-i\delta\theta
    \left\{\bfvtheta_{+,0}{\rm C}+\bfvtheta_{-,0}\right\}A(\bfx),
    \label{gauge-fin}
    \ee
where $\bfvtheta_{\ep,0}:\cHa\to\cHa$ is defined by
    \be
    \bfvtheta_{\ep,0}:=[L^+_\ep-L^0_\ep]\fH^2+L^-_\ep\fH+L^0_\ep.
    \label{tTheta-ep0}
    \ee

In view of (\ref{gauge-fin}), the symmetry transformations
(\ref{trans}) are generated by the operator $\bfvtheta_{+,0}{\rm
C}+\bfvtheta_{-,0}$. We can easily exponentiate the latter to
obtain the corresponding non-infinitesimal symmetry
transformations,
    \be
    A\to e^{-i\theta\left(\bfvtheta_{+,0}{\rm C}+\bfvtheta_{-,0}\right)}A,
    \label{gauge-exp}
    \ee
where $\theta\in\R$ is arbitrary. In terms of the
$(\ep,h)$-components $A_{\ep,h}$ of $A$, (\ref{gauge-exp}) takes
the form
    \be
    A=\sum_{\ep=\pm}\sum_{h=0,\pm1}A_{\ep,h}
    \to \sum_{\ep=\pm}\sum_{h=0,\pm1}e^{-i\ep\fa_{\ep,h}}A_{\ep,h},
    \label{gauge-eph-comp}
    \ee
where we have made use of ${\rm C}^2=1$ and $\fH^3=\fH$.

Similarly to its spin-zero counterpart \cite{MZ-ann06-1}, as seen
from (\ref{gauge-exp}) and (\ref{gauge-eph-comp}), the gauge
group\footnote{Here we identify the gauge group with its connected
component that includes the identity and is obtained by
exponentiating the generator $\bfvtheta_{+,0}{\rm
C}+\bfvtheta_{-,0}$.} $G\suba$ associated with these
transformations is a one-dimensional connected Abelian Lie group.
Therefore, it is isomorphic to either of $U(1)$ or $\R^+$, the
latter being the noncompact multiplicative group of positive real
numbers \cite{brocker-dieck}.

We can construct a faithful representation of the group $G\suba$
using the six-component representation
$A=\left(A_{+,+1},A_{+,-1},A_{+,0},A_{-,+1},A_{-,-1},A_{-,0}
\right)^T$ where ${\rm C}$ and $\fH$ are, respectively,
represented by $\Sigma_3$ and $\Sigma_{12}$. In this
representation a typical element of $G\suba$ takes the form
    \be
    g_a(\theta):={\rm diag}\left(
    e^{-i\fa_{+,+1}\theta},e^{-i\fa_{+,-1}\theta},e^{-i\fa_{+,0}\theta},
    e^{i\fa_{-,+1}\theta},e^{i\fa_{-,-1}\theta},e^{i\fa_{-,0}\theta}
    \right).
    \label{element}
    \ee
This expression suggests that the gauge group $G\suba$ is a
subgroup of $U(1)\otimes^{6}$, the latter being the direct product
of six copies of $U(1)$. It is not difficult to show that $G\suba$
is a compact subgroup of this group and consequently isomorphic to
$U(1)$ if and only if all the parameters $\fa_{\ep,h}$ are
rational numbers, otherwise $G\suba$ is isomorphic to $\R^+$.

Clearly, the $G\suba$ gauge symmetry associated with the
conservation of the total probability is a global gauge symmetry.
Similarly to its spin-zero counterpart \cite{MZ-ann06-1} the local
analog of this global gauge symmetry is different from the usual
local Yang-Mills-type gauge symmetries.

\section{Conclusion and Discussion}

In this article we have used the methods of pseudo-Hermitian
quantum mechanics to devise a complete formulation of the
relativistic quantum mechanics of the Proca fields that does not
involve restricting to the positive-energy solutions of the Proca
equation. In particular, we have constructed the most general
physically admissible inner product $\cbr\cdot,\cdot\para$ on the
solution space of the Proca equation. Up to a trivial scaling,
this inner product involves five real parameters that we
collectively denote by $\fa$. For all the values of $\fa$ the
inner product $\cbr\cdot,\cdot\para$ is positive-definite and
relativistically invariant. It also renders the generator of the
time-translations, i.e., the Hamiltonian ${\rm h}$, and the
helicity operator self-adjoint. The quantum system associated with
the Proca fields may be represented by a Hilbert space $\cHa$
defined by the inner product $\cbr\cdot,\cdot\para$ and the
Hamiltonian ${\rm h}$. Using the unitary equivalence of this
covariant representation with that of the noncovariant
Foldy-representation, which arises quite naturally in our
formulation, we constructed relativistic position, momentum,
angular momentum, spin and helicity operators acting in $\cHa$,
and the localized states of the Proca field. Furthermore, we
introduced the position wave functions and used them to construct
a probability current density which turned out to be neither
conserved nor covariant. We resolved the apparent inconsistency of
this observation with the physical requirement of the conservation
and frame-independence of the total probability using a conserved
and covariant current density. The global conservation of the
total probability is supported by the local conservation of this
current density. The latter is linked to a previously unnoticed
global gauge symmetry of the Proca field with an Abelian gauge
group. In Ref.~\cite{Jakubsky}, the authors have also attempted to
give a spin-one generalization of our treatment of the
Klein-Gordon fields. However, because of various self-imposed
restrictions they obtain a one-parameter subfamily of the inner
products $\cbr\cdot,\cdot\para$.\footnote{The treatment of
\cite{Jakubsky} has some minor errors. For example, the system
$\{\Psi_{\ep,h}(\bfk),\Phi_{\ep,h}(\bfk)\}$ used in
\cite{Jakubsky} to construct the metric operator is actually not
biorthonormal, as claimed by the authors. Fortunately, this type
of errors could be absorbed in the unknown parameters of the
metric operator and do not affect the final result.} This is
related to the fact that the authors of \cite{Jakubsky} use the
special Foldy transformation (\ref{u-A=}) (see Eq.~(35) of
\cite{Jakubsky}) to find the metric operator acting in the Hilbert
space $\cH'$. Furthermore, in trying to impose the condition of
the Lorentz-invariance of the metric operator, they transform the
operator $\teta_+$ to a metric operator acting in the Foldy
representation and demand that the latter commutes with the
generators of the Poincar\'{e} group in this representation. This
does not seem to be well-justified, because as seen from
(\ref{inn-p-6}), the Hilbert space of the Foldy representation is
just the direct sum of two copies of $\tcH$. Hence, the metric
operator associated with this representation is just the identity
operator, not the one given by Eq.~(39) of \cite{Jakubsky}. As
seen from (\ref{tu-A=}), this is the Foldy transformation which
depends on the choice of the unknown parameters appearing in the
operator $\teta_+$, not the metric operator of the Foldy
representation. We would also like to stress that the analysis of
\cite{Jakubsky} does not include the construction of the
observables of the system or any treatment of its physical aspects
such as the notorious problem of the probabilistic interpretation
of the quantum mechanics of Proca fields.

Apart from the historical importance of the subject, the present
work is mainly motivated by the close analogy of the Proca and
Maxwell fields. Performing the zero-mass limit of our results in
an appropriate manner should lead to a consistent quantum
mechanical treatment of individual photons. As it is to be
expected, there are subtleties in performing this limit.
Nevertheless, we have been able to make some progress toward
solving the problem of the construction of the Hilbert space and
observables for the photon. We plan to report the results in a
separate article.

\section*{Acknowledgment}
F.~Z.~wishes to acknowledge the hospitality of Ko\c{c} university
during his visits.


\ed